%% file: main.tex
\documentclass[sigconf,nonacm]{acmart}

\usepackage{amsmath}
\usepackage{amssymb,amsfonts}
\usepackage{textcomp}
\usepackage{xcolor}
\usepackage{url}
\usepackage{hyperref}
\usepackage{seqsplit}

\definecolor{yellow}{RGB}{255,255,153}
\definecolor{grey}{RGB}{224,224,224}
\definecolor{green}{RGB}{0,100,0}

\usepackage{algorithm}
\usepackage{algorithmic}
\usepackage{graphicx}
\usepackage{subfigure} 
\usepackage{fancybox}
\usepackage{xspace}
\usepackage{threeparttable}
\usepackage{multirow}
\usepackage{enumitem}
\usepackage{seqsplit}
\usepackage{pifont}
\usepackage{caption}

\newcommand{\framework}{\textsc{IDRadar}}

\newcommand{\datasetsize}{1,814 }
\newcommand{\totalbrand}{250 }
\newcommand{\totalsensitiveproperties}{272 }
\newcommand{\totalsensitivepropertiescase}{8,192 }
\newcommand{\totalsensitivesettings}{104 }
\newcommand{\totalsensitivesettingscase}{3,620 }

\newcommand{\totaldangerouspropertiescase}{3,477 }

\newcommand{\totaldangeroussettingscase}{1,336 }
\newcommand{\totalsensitivedevices}{1,406 }
\newcommand{\totaldangerousdevices}{1,112 }

\newcommand{\dynamicmodels}{314 }
\newcommand{\dynamicbrands}{22 }

\newcommand{\dynamicpropertiescase}{160 }

\newcommand{\dynamicsettingscase}{400 }

\newcommand{\dynamictotal}{216 }

\begin{document}
\title{On the (In)Security of Non-resettable Device Identifiers in Custom Android Systems}

\author{Zikan Dong}
\affiliation{%
  \institution{Beijing University of Posts and Telecommunications}
  \city{Beijing}           
  \country{China}
}

\author{Liu Wang}
\affiliation{%
  \institution{Beijing University of Posts and Telecommunications}
  \city{Beijing}           
  \country{China}
}

\author{Guoai Xu}
\affiliation{%
  \institution{Harbin Institute of Technology, Shenzhen}
  \city{Shenzhen}           
  \country{China}
}

\author{Haoyu Wang}
\authornotemark[2]
\affiliation{%
  \institution{Huazhong University of Science and Technology}
  \city{Wuhan}           
  \country{China}
}

\begin{abstract}
User tracking is critical in the mobile ecosystem, which relies on device identifiers to build clear user profiles. 
In earlier ages, Android allowed easy access to non-resettable device identifiers like device serial numbers and IMEI by third-party apps for user tracking.
As privacy concerns grew, Google has tightened restrictions on these identifiers in native Android.
Despite this, stakeholders in custom Android systems seek consistent and stable user tracking capabilities across different system and device models, and they have introduced covert channels (e.g., system properties and settings) in customized systems to access identifiers, which undoubtedly increases the risk of user privacy breaches.
This paper examines the introduction of non-resettable identifiers through system customization and their vulnerability due to poor access control.
We present {\framework}, a scalable and accurate approach for identifying vulnerable properties and settings on custom Android ROMs.
Applying our approach to \datasetsize custom ROMs, we have identified \totalsensitivepropertiescase system properties and \totalsensitivesettingscase settings that store non-resettable identifiers, with \totaldangerouspropertiescase properties and \totaldangeroussettingscase settings lacking adequate access control, which can be abused by third-party apps to track users without permissions. Our large-scale analysis can identify a large number of security issues which are two orders of magnitude greater than existing techniques.
We further investigate the root causes of these access control deficiencies.
Validation on 32 devices through the remote real-device testing service confirmed our results.
Additionally, we observe that the vulnerable properties and settings usually occur in devices of the same OEMs. 
We have reported our findings to the respective vendors and received positive confirmations.
Our work underscores the need for greater scrutiny of covert access channels to device identifiers and better solutions to safeguard user privacy during system customizations.

\end{abstract}

\maketitle

\input{Sections/introduction}
\input{Sections/background}
\input{Sections/experiment}
\input{Sections/results}
\input{Sections/discussion}
\input{Sections/related.tex}
\input{Sections/conclusion}


\newpage
\bibliographystyle{ACM-Reference-Format}
\bibliography{base}
\input{Sections/appendices}

\end{document}

%% file: Sections/introduction.tex
\section{Introduction}
\label{sec:introduction}

User tracking is pivotal in the Android ecosystem, enabling apps to deliver personalized user experiences, targeted advertising, user behavior analysis, performance monitoring, and market segmentation~\cite{kollnig2022iphones, binns2018third, razaghpanah2018apps, lerner2016internet, nath2015madscope, yang2020comparative}. 
This practice allows developers to optimize app functionality, enhance marketing strategies, and deliver tailored content to users, thereby significantly contributing to the overall effectiveness and user satisfaction of mobile apps.
In Android, identifiers, which are unique strings or numbers, are crucial for tracking by distinguishing users or devices and are often collected with specific user data.
The uniqueness of identifiers allows apps to link behaviors collected at different times, locations, and even across different apps to the same user, enabling them to build a clear user profile and provide better services.

Android identifiers can be classified based on their scope of usage. 
\textit{App identifiers} are unique to each app and are used to track user behavior within a single app, while \textit{Device identifiers} are used to track user behavior across multiple apps on the same device. 
Device identifiers can be further categorized based on their persistence into \textit{user-resettable} and \textit{non-resettable} types. 
User-resettable identifiers, like advertising IDs~\cite{advertisingid}, can be reset by the user, providing control over privacy. 
In contrast, non-resettable identifiers, such as the International Mobile Equipment Identity (IMEI), are permanently tied to the device and provide continuous tracking even after a device reset, making them preferable for tracking, but their high stability and extensive reach come with significant privacy concerns.
If compromised, these identifiers could lead to severe breaches of user privacy, potentially exposing extensive personal data to malicious actors.
Additionally, non-resettable device identifiers are widely used for user tracking not only by third-party apps, but also by various supply chain actors in the Android ecosystem, such as OS developers, hardware vendors, and pre-installed system apps~\cite{lyons2023log, gamba2020analysis, schindler2022privacy, leith2021mobile, liu2021android, reardon201950, demetriou2016free, chen2014information}.
This widespread use highlights the essential requirement for robust protective measures to prevent their compromise and ensure user privacy.

As privacy concerns grow, the management of non-resettable device identifiers in Android has undergone significant updates to improve user privacy.
Before Android 10, apps could access these identifiers through specific official APIs after obtaining the required permissions and user consent.
With the release of Android 10 in 2019, access to these identifiers was restricted to apps with privileged permissions, which are not available to third-party apps.
Google now recommends third-party apps use user-resettable advertising IDs for advertising and analytics purposes.
Concurrently, many countries have enacted legislation (e.g., GDPR~\cite{gdpr} and CCPA~\cite{ccpa}) to regulate the collection and use of personal data.

However, some supply-chain actors, including device and hardware vendors as well as software providers, often introduce \textit{covert channels} to facilitate the retrieval of non-resettable identifiers~\cite{meng2023post, zhou2022uncovering}.
For instance, storing these identifiers in \textit{system properties and settings}, which function as global variables in the Android system, allows easy access to them.
One possible reason for this could be that these supply chain actors desire to maintain consistent, stable, and convenient tracking capabilities across different versions or modifications of the system. 
Introducing covert access channels can enable them to access these stable device identifiers, regardless of changes made to the official APIs or other system components by Google and system customizers.
However, this practice raises privacy concerns, as non-resettable identifiers could be leaked through these channels, creating new attack surfaces, if not properly managed.
These channels should be secured with access control measures equivalent to those protecting the identifiers themselves, necessitating stringent access control policies during system customization.
Regrettably, the customization introduced by vendors tends to introduce vulnerabilities such as improper security configurations and outdated protections, as highlighted in many existing studies~\cite{possemato2021trust, hou2022large, elsabagh2020firmscope, liu2022customized, el2021dissecting, li2021android}.
Inadequate protection of these identifiers can lead to unauthorized user tracking, impacting user privacy.

The security of custom Android systems has been widely studied by the research community.
Many researchers have focused on pre-installed apps in custom systems, studying privacy concerns or additional attack surfaces introduced by these apps~\cite{blazquez2021trouble, gamba2020analysis, elsabagh2020firmscope}.
There has also been research into the timeliness and effectiveness of applying Google's security updates in custom environments~\cite{hou2022large, liu2022customized, el2021dissecting}.
These studies, although have touched on the privacy issues within custom Android systems, they primarily consider the official channels for user tracking and do not take into account the covert channels that can return non-resettable device identifiers, introduced by the system customizations.
To the best of our knowledge, the only existing work in the research community that mentions these channels is U2-I2~\cite{meng2023post}, which employs dynamic testing on real devices to assess the mismanagement of these channels. 
However, their approach limits its research breadth.
U2-I2 requires a factory reset of devices, which confines them to using physical devices (even cannot use the widely accessible remote real-devices as alternatives). 
Consequently, expanding the scope of experiments necessitates purchasing a large number of devices, making it cost-prohibitive. 
Additionally, their method involves considerable manual costs, which also limits the test coverage.
As a result, it covered only 13 smartphones from 9 vendors, identifying 23 vulnerable system properties and 7 settings. 

In the work, we aim to provide a large-scale, accurate, and expandable investigation of non-resettable device identifiers in the wild, including the covert channels (i.e., system properties and settings), their usage, and the involved vulnerabilities.
To this end, we present {\framework}, an end-to-end approach that proceeds in three steps.
Firstly, 
we target the usage of system properties and settings in custom ROMs\footnote{An Android ROM is a firmware image file that contains the operating system and software for Android devices, enabling users to install or update on their devices.}
to identify all relevant instances in custom Android systems.
This begins with a pilot study to catalog methods of accessing system properties and settings in custom systems, followed by a static analysis method that leverages control-flow and data-flow analysis to pinpoint their usage points.
Secondly, we filter the above results by employing two heuristic approaches to identify potential system property and setting candidates that store non-resettable device identifiers.
We further manually verify the content of these properties and settings through the analysis of contextual code information.
Finally, we examine SELinux policies and system frameworks in custom systems to identify the access control for system properties and settings, uncovering vulnerable ones that store non-resettable device identifiers without adequate access control, i.e., can be abused by third-party apps without permissions.

Applying our methodology to \datasetsize custom Android systems from \totalbrand different Original Equipment Manufacturers (OEMs), 
we identified \totalsensitivepropertiescase system properties and \totalsensitivesettingscase system settings that store non-resettable device identifiers. 
Among these, \totaldangerouspropertiescase system properties and \totaldangeroussettingscase system settings across \totaldangerousdevices custom Android systems were found to lack proper access control measures.
Our detailed analysis of these vulnerable cases revealed major reasons for such vulnerabilities, including overlooked system properties and settings, overly complex access control rules, and additional access channels for system properties. 
To validate our results, we developed a demo app that requires no permissions and can access the device identifiers stored in these system properties and settings. We tested this app on 32 devices available through remote real-device testing services, and identified 130 vulnerable system properties and settings.
Of them, 102 can be directly triggered on these real devices and the remaining 28 have specific trigger conditions, confirming the accuracy of our detected results.
We further conducted a detailed comparison with U2-I2~\cite{meng2023post} on four real Android devices. The result shows that we can indeed identify more identifiers in systems properties and settings, which highlights U2-I2’s limitation in detecting issues that rely on specific preconditions (e.g., activating a feature or performing an action), whereas our static approach is capable of performing system-wide comprehensive analysis.
We also observed that vulnerable implementations usually recur across devices from the same OEMs. We then extended our research to a further \dynamicmodels real devices of \dynamicbrands OEMs using the same testing service. 
Even without analyzing these specific systems, we were able to directly trigger \dynamicpropertiescase vulnerable system properties and \dynamicsettingscase vulnerable system settings that store non-resettable device identifiers, affecting a total of \dynamictotal devices. 
We have reported the identified issues to the respective device vendors, and 84 vulnerabilities have been confirmed as of this writing. 

In summary, our work presents these major contributions:

\begin{itemize}
    \item We conduct a large scale study on custom Android systems to examine covert channels (i.e., custom system properties and settings) for accessing non-resettable device identifiers. 
    Our study presents an overall landscape of the privacy issues introduced by these covert channels.

    \item We develop an end-to-end analysis pipeline to thoroughly investigate the covert channels on a large dataset of custom Android ROMs, successfully identifying thousands of system properties and settings that store non-resettable device identifiers, many of which are vulnerable to exploitation by third-party apps. We also analyze the root cause of these vulnerabilities.

    \item We leverage remote real-device testing services to validate our approach, demonstrating that it is possible for third-party apps to retrieve non-resettable device identifiers from real devices without requesting any permissions. We have reported these to the respective vendors and receive positive confirmation.
    
\end{itemize}

The source code of our tool is publicly available~\cite{opensource}.

%% file: Sections/background.tex
\section{Background}
\label{sec:background}

\subsection{Non-resettable Device Identifiers}
\label{sec:background0}

\begin{table*}[ht]
\caption{SELinux Key Concepts Explanation.}
\label{tab:selinux}
\resizebox{1\textwidth}{!}{
\begin{tabular}{|l|l|l|}
\hline
SELinux Key   Concepts         & Format                                                                              & Example                                                                                                       \\ \hline
Property Context                & property\_name user:role:{\color{red}type}:sensitivity{[}:categories{]}             & \textbf{(a)} xxx.xxx.xxx.imei1 u:object\_r:{\color{red}system\_id\_prop}:s0                             \\ \hline
\multirow{3}{*}{Policy Rule}   & \multirow{2}{*}{allow {\color{blue}source\_type} {\color{red}target\_type}:class permissions}   & \textbf{(b)} allow {\color{blue}system\_app} {\color{red}system\_id\_prop} (file (read getattr map open))     \\ \cline{3-3} 
                               &                                                                                     & \textbf{(c)} allow {\color{blue}radio} {\color{red}system\_id\_prop} (property\_service (set))                \\ \cline{2-3} 
                               & allow type\_attribute\_set type\_attribute\_set:class permissions                   & \textbf{(d)} allow appdomain extended\_core\_property\_type (file (read getattr map open)))                                \\ \hline
Type Attribute Set             & typeattributeset {\color{blue}attribute\_name} {[}(and {\color{red}type1} ...){]} {[}(not {\color{green}type1} ... ){]} & \textbf{(e)} typeattributeset {\color{blue}extended\_core\_property\_type} ({\color{red}system\_id\_prop} ... ) \\ \hline
Keyword expandtypeattribute    & (expandtypeattribute attribute\_name $<$true, false$>$)                                  & \textbf{(f)} (expandtypeattribute extended\_core\_property\_type true)                                         \\ \hline
\end{tabular}
}
\end{table*}

In Android, non-resettable device identifiers are unique identifiers that cannot be altered or reset by the user, providing a consistent means of identifying a device. 
In our study, we consider the seven most commonly used non-resettable device identifiers, which are officially documented and frequently referenced in numerous researches~\cite{meng2023post, razaghpanah2018apps, reardon201950, ren2018longitudinal, stevens2012investigating}.
\begin{itemize}
    \item[1)] International Mobile Equipment Identity (IMEI): a fixed-length decimal digits assigned to mobile devices used for identifying valid devices on a cellular network. 
    \item[2)] Mobile Equipment Identifier (MEID): similar to the IMEI but used in CDMA phones.
    \item[3)] International Mobile Subscriber Identity (IMSI): a unique identifier assigned to the user of a cellular network, stored on the SIM card, often used for authenticating and tracking users on mobile networks.
    \item[4)] Integrated Circuit Card Identifier (ICCID): a unique identifier for a SIM card itself, used for identifying the card internationally and managing its activation and deactivation. Unlike the IMSI, which identifies the user, the ICCID is used to identify the SIM card.
    \item[5)] Device serial number: a unique identifier assigned by the device manufacturer, used for device warranty and service management. 
    \item[6)] WiFi MAC address: a unique identifier for the network interface card of a device, used for network access control and tracking devices on WiFi networks. 
    \item[7)] Bluetooth MAC address: a unique identifier for the Bluetooth module of a device. 
\end{itemize}

These identifiers are crucial for various functions, such as device tracking, network access, and device management.
The broad scope and stable nature of these identifiers are essential for user tracking, but they also pose significant privacy risks. Thus, they are no longer permitted for use by third-party apps after Android 10.

\subsection{System Properties and Settings}
\label{sec:background1}
\textit{\textbf{System properties}} in Android are key/value pairs used by system components and apps to manage configuration parameters~\cite{system_properties}.
These properties are managed by the property service, which sets up a shared memory area to store them during system startup. 
Some system properties are loaded from configuration files like build.prop and default.prop, which store fixed information (e.g., device name and system version).
Additionally, some system properties are dynamically set at runtime, such as device-specific information like the serial number, or runtime status of the device.
Accessing system properties simply requires querying the shared memory using system APIs or commands, without requiring interprocess communication.
And, setting properties involves direct communication with the property service, which handles the updates. 
This implementation makes system properties easy to use and ensures they have minimal performance overhead.

\textit{\textbf{System settings}} are also key/value pairs used to store device preferences, accessible system-wide, and can be easily managed through system APIs.
System settings are primarily stored in databases, and divided into three categories including \textit{System}, \textit{Secure}, and \textit{Global} settings.
Each category of system setting has a different purpose: \textit{Global settings} apply uniformly across all users, \textit{Secure settings} are readable but not writable by apps, and \textit{System settings} store miscellaneous system preferences~\cite{system_settings}.
The settings are managed by the \texttt{SettingsProvider}, which is a content provider in the Android system. 
It handles read and write operations to the database, ensuring data consistency and security.

\subsection{Access Control Policies}
\label{sec:background2}

The scope of system properties and settings is system-wide, allowing all entities within the system, such as system services, system apps, and third-party apps, to access.
However, since they function like global variables and store substantial device information, insufficient access control can result in vulnerabilities and exploitation.
For example, system properties and settings storing configuration parameters should be set only by the responsible system components, while those holding sensitive device information should not be accessible to apps without appropriate permissions.

In Android, Security-Enhanced Linux (SELinux)~\cite{selinux} enforces access control for system properties by implementing mandatory access control (MAC)~\cite{mac}.
SELinux operates on a principle of default denial, where any action not explicitly permitted will be denied.
Specifically, SELinux uses ``property contexts'' (a type of security context) to assign type attributes to properties, as well as processes within the system.
And specific access control permissions are assigned to these types.
The format of property contexts is shown in~\autoref{tab:selinux} example (a).
After the type attribute of a system property is identified, SELinux determines which behaviors are allowed based on specific policy rules.
For example, the example (b) in~\autoref{tab:selinux} shows that the system app (source type) is allowed to read system properties of type “system\_id\_prop” (target type).
As discussed in~\S\ref{sec:background1}, accessing properties involves reading from files representing shared memory. 
Thus, read permissions to these files enable objects to access system properties.
For setting properties, it is necessary to grant objects the ability to set the property service, as shown in example (c) of~\autoref{tab:selinux}.
In terms of format, these SELinux rules exist in ``te'' files (Type Enforcement) before system compilation. 
When compiled and packaged into the system ROM, these rules are converted into binary files. 
With the introduction of Project Treble~\cite{treble} in Android 8.0, a new intermediate language Common Intermediate Language (CIL) is used to define SELinux rules.
Other SELinux-related concepts are explained in appendices.

Besides, Android also has access control policies for system settings.
On the one hand, writing system settings is mainly controlled by permissions, such as the \texttt{\seqsplit{WRITE\_SETTINGS}}~\cite{write_settings} and \texttt{\seqsplit{WRITE\_SECURE\_SETTINGS}}~\cite{write_secure_settings} permissions.
Only apps with the appropriate permissions can add or change system settings, and the \texttt{\seqsplit{WRITE\_SECURE\_SETTINGS}} permission can only be used by system apps.
In terms of reading system settings, before Android 12, only settings labeled as ``@SystemApi'' were not accessible to third-party apps while all other settings were accessible.
Starting from Android 12, the ``@Readable'' annotation is introduced in the Android Open Source Project (AOSP)~\cite{aosp} to refine access control.
Specifically, the system rejects access to settings from non-system apps attempting to access settings defined in Settings.Secure, Settings.System, or Settings.Global that lack the ``@Readable'' annotation.
If a system setting has both the ``@Readable'' and ``@SystemApi'' annotations, only system apps can access it.
Furthermore, for system settings not explicitly defined in the Settings class but added by privileged apps, we observed that there is no access control mechanism in place, affecting both versions before and after Android 12.

%% file: Sections/experiment.tex
\section{{\framework}}
\label{sec:approach}
To thoroughly understand the covert access channels (i.e., custom system properties and settings) for non-resettable device identifiers in large-scale custom Android systems, we design {\framework}, which will be depicted in the following.

\begin{figure*}[ht]
  \centering
  \resizebox{1\textwidth}{!}{
  \includegraphics[width=\linewidth]{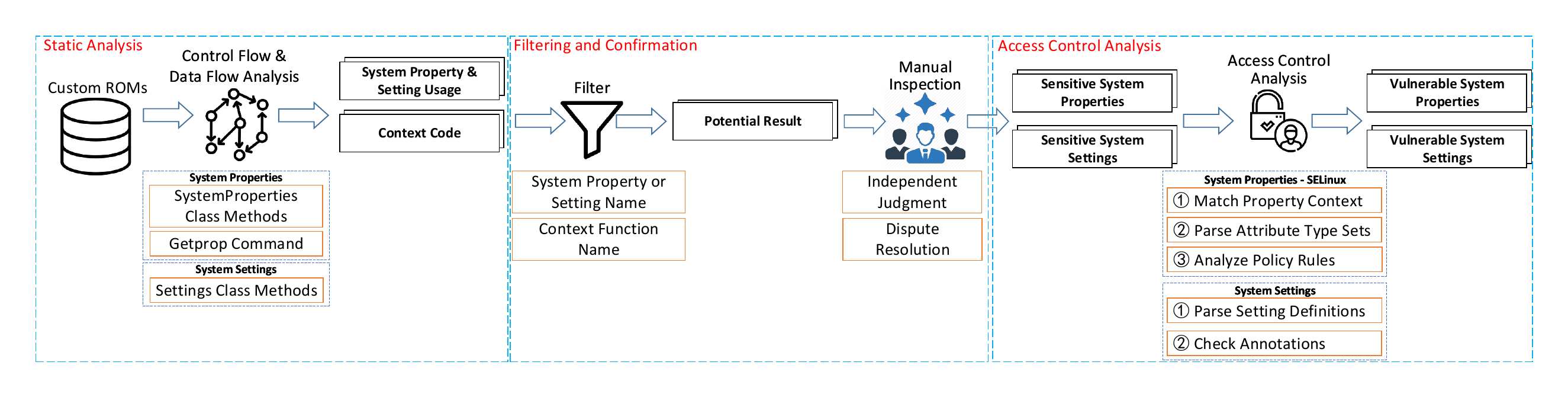}
  }
  \caption{The Overview of {\framework}.}
  \label{fig:approach}
\end{figure*}

\subsection{Key Ideas and Pilot Study}
\label{sec:pilot}

To study the custom system properties/settings that store non-resettable device identifiers, one straightforward idea is to identify all locations where system properties or system settings are defined and then analyze their content.
In practice, unfortunately, finding these definitions is challenging, as they may be located across multiple layers of the Android system (e.g., the Linux kernel, drivers, system libraries, frameworks, pre-installed apps), and lack a consistent definition format.
Therefore, we take a different approach -- 
we focus on the usage of these system properties and settings, and analyze their content based on contextual information.
The assumption of this method is that the custom system properties and settings storing device identifiers are introduced to serve the functions of the system customization parts. 
And, the probability that the system defines a system property or system setting without ever using it is low.
Hence, we argue that locating the usage of system properties and system settings, and determining their content, can still provide a sufficient understanding of the custom system properties and settings.
We admit that this assumption is difficult to validate, as the ground truth of all custom system properties and settings is hard to obtain, which will be discussed in the limitation.

Although finding usage is a relatively feasible method, it is not that trivial, especially when it comes to locating the usage of system properties.
Unlike system settings, which can be accessed through well-documented methods, system properties are often accessed via ``non-SDK methods\footnote{Non-SDK methods encompass those not included in the public SDK, often belonging to internal APIs or used only in system-level apps only, leading to variable access behaviors.}'' or system commands.
Therefore, the behavior of accessing system properties is more variable.
To comprehensively identify the usage points of system properties, we first conducted a pilot study to investigate methods of accessing system properties.

In the pilot study, we attempt to identify potential system property access behaviors.
Since system property names typically appear as strings with format characteristics like starting with ``ro.'', ``persist.'', or ``vendor.''~\cite{meng2023post}.
We extract DEX files from APK and JAR files in the custom system and use the ``strings'' command to search for property names and further determine the presence of property access behaviors. 
Then, we manually reverse-engineer these files and summarize the methods they use to access system properties.
This pilot study enable us to understand the diverse ways system properties are accessed in custom systems.
The results of the pilot study will be explained in detail in~\S\ref{sec:approachstatic}.

\subsection{Overview of {\framework}}
Our methodology is depicted in~\autoref{fig:approach}.
We begin with a static analysis of all APK and JAR files within the custom Android ROMs.
Leveraging the insights from our pilot study, we pinpoint all instances of system property and setting usage, along with their contextual code.
Next, we apply a heuristic algorithm to filter the extensive results from the first step, identifying a subset likely to contain non-resettable device identifiers as mentioned in~\S\ref{sec:background0}.
Security experts then manually review the filtered subset to confirm which properties or settings store sensitive identifiers.
Finally, we analyze SELinux policies and framework code in custom systems to determine the access control policies of these system properties and settings, identify the vulnerable ones.

\subsection{Static Analysis}
\label{sec:approachstatic}

\subsubsection{Methods of accessing system properties/settings}
\begin{figure}[h]
  \centering
  \resizebox{1\linewidth}{!}{
  \includegraphics[width=\linewidth]{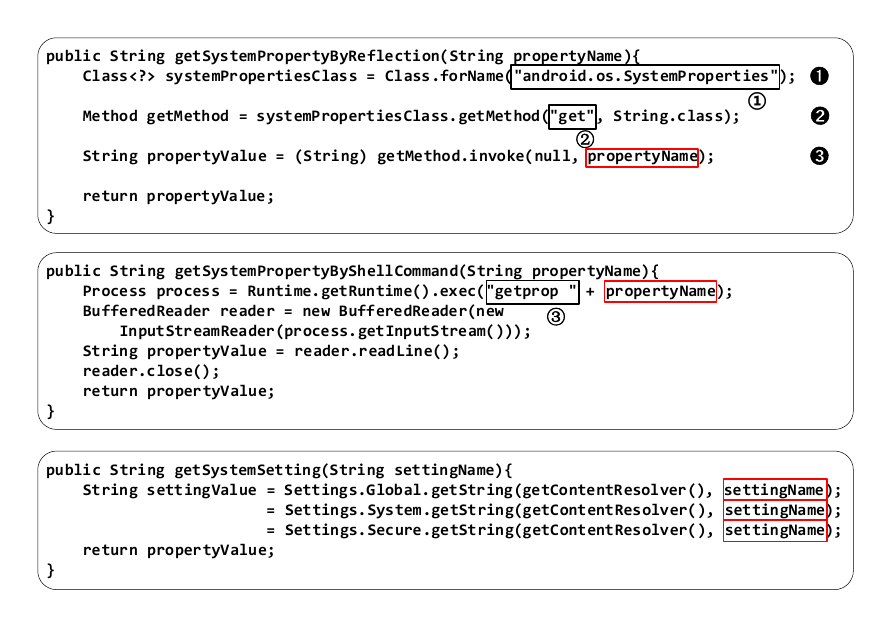}
  }
  \caption{Code for Accessing System Properties and Settings.}
  \label{fig:code}
\end{figure}

Through the pilot study in~\S\ref{sec:pilot}, we summarize the methods for accessing system properties and settings.
System properties is primarily accessed through non-SDK methods in the \texttt{\seqsplit{android.os.SystemProperties}} class or using the \texttt{getprop} system command.
For the \texttt{getprop} command, system code typically executes system commands using the \texttt{exec} method of the \texttt{\seqsplit{java.lang.Runtime}} class, as shown in~\autoref{fig:code}.
For the methods of the \texttt{\seqsplit{SystemProperties}} class, only a small amount of system code can directly call these methods.
Most code requires invocation through Java reflection. 
Typically, reflection involves at least three steps, as shown in~\autoref{fig:code}.

In this code snippet, the process begins by dynamically loading the \texttt{\seqsplit{android.os.SystemProperties}} class using the \texttt{\seqsplit{Class.forName}} method (Step~\ding{202}). 
Next, the ``get'' method of the SystemProperties class is retrieved via \texttt{\seqsplit{Class.getMethod}} method (Step~\ding{203}). 
The method is then invoked with \texttt{\seqsplit{Method.invoke}} method, where ``propertyName'' is the specific system property name to access (Step~\ding{204}). 
The return value of this invocation is a String, which is the value of the requested system property.
However, through our pilot study, we found that there are different coding styles for reflection invocations.
All approaches to calling methods in \texttt{\seqsplit{SystemProperties}} class can be summarized as follows:
(a) Directly invoking methods in the SystemProperties class without using reflection. 
(Some system components can use this approach.)
(b) Using reflection, with all three steps of reflection contained within a single method.
(c) Using reflection, but the first two steps are completed during the static initialization of the calling class. 
The \texttt{\seqsplit{Method}} object obtained in the second step is stored in a static field, allowing the third step to be executed directly using this static field without reinitialization.
(d) Using reflection, but the first step or the first two steps are wrapped within a method. 
Subsequently, the wrapper method is called to obtain the ``Class'' or ``Method'' object.
Besides, for system settings, Android provides related SDK APIs. 
Most system code uses methods provided by the \texttt{\seqsplit{android.provider.Settings}} class to access system settings.

\subsubsection{Identifying system properties/settings}

After determining the methods of accessing system properties and settings, we aim to identify all the usage of system properties/settings in the custom ROMs, along with the contextual code that can help determine their content.
We employ static analysis, built upon Soot~\cite{vallee2010soot}, to analyze all APK and JAR files in the custom ROMs.
We first decompile these files and convert the Java code into Jimple, an intermediate representation used in Soot.
We then traverse all methods to locate those for accessing properties/settings, focusing primarily on invoke expressions and assignment statements with field references.

For invoke expressions, on the one hand, we examine whether the called methods are related to accessing system properties and settings, including methods in the \texttt{\seqsplit{android.os.SystemProperties}} class for accessing system properties, the \texttt{\seqsplit{java.lang.Class: forName}} method for Java reflection, the \texttt{\seqsplit{java.lang.Runtime: exec}} method for executing system commands, and methods in the \texttt{\seqsplit{android.provider.Settings}} class for accessing system settings.
On the other hand, we focus on whether the return type of a function call is \texttt{\seqsplit{java.lang.Class}} or \texttt{\seqsplit{java.lang.reflect.Method}}, corresponding to type (d) mentioned above for accessing system properties.
For assignment statements containing field references, we focus on whether the field type is \texttt{\seqsplit{java.lang.reflect.Method}}, corresponding to type (c) for accessing system properties.

When dealing with Java reflection and system commands execution through the \texttt{\seqsplit{java.lang.Runtime: exec}} method, we further analyze the content of the method parameters (i.e., \ding{202}, \ding{203}, and \ding{204} in~\autoref{fig:code}) to verify whether the reflected methods or executed commands are pertain to system properties and settings access.
This requires performing inter-procedural analysis. 
First, to handle types (c) and (d) for accessing system properties, when we find (1) a field type or (2) a method return type is ``Class'' or ``Method'', we further analyze the initialization of the class containing the field for situation (1); the methods returning ``Class'' or ``Method'' objects, to confirm whether the reflected call is to a \texttt{\seqsplit{SystemProperties}} class method for situation (2).
Second, to handle the instance of method encapsulation, where the reflected method value or executed system commands come from method parameters. 
We use a generated Inter-procedural Control Flow Graph (ICFG) to locate the call sites of the method and confirm the corresponding parameter values.

\subsubsection{Analyzing system properties/settings}
Next, for the confirmed system properties and settings access behaviors, we analyze the names of the accessed system properties/settings. 
These names serve as parameters in key methods, such as ``propertyName'' and ``settingName'' in~\autoref{fig:code}.
During this process, some special cases require additional handling: (1) The property or setting name is passed as a parameter. 
In this case, we conduct inter-procedural analysis to determine the parameter's value; 
(2) The name is constructed through string concatenation. 
For this, we simulate the string operations to retrieve the full name; 
(3) The name is stored in a field, where we look for the name in the class's \texttt{\seqsplit{init}} method; 
(4) The name is placed in an array, and we extract all the contents of the array, treating them as potential names.

Once we have obtained the names of the system properties and settings accessed in the custom system, we extract the context code of these usage points to analyze their content. 
To do this, we use the ICFG to backtrack from the call points (identifying the callers of the method), obtaining the complete method call chain and code content as contextual information.

\subsection{Filtering and Confirmation}
\label{sec:approachfilter}
Through static analysis, we obtained numerous system properties and settings used in the custom ROM, along with the corresponding context code.
Next, we aim to pinpoint a subset of these system properties and settings that contain the seven types of non-resettable device identifiers outlined in~\S\ref{sec:background0}.
To achieve this, we implement two simple yet effective heuristic algorithms: (1) We select system properties and settings whose names include keywords related to device identifiers, such as IMEI, ICCID, DeviceID, etc. 
In custom systems, these names generally reflect their content, without obfuscation.
(2) For context method whose name contains keywords related to device identifiers, e.g., ``getIMEI'' and ``getWifiMac'',we infer that the system properties/settings accessed within these methods correspond to those identifiers.

\begin{figure}[h]
  \centering
  \resizebox{1\linewidth}{!}{
  \includegraphics[width=\linewidth]{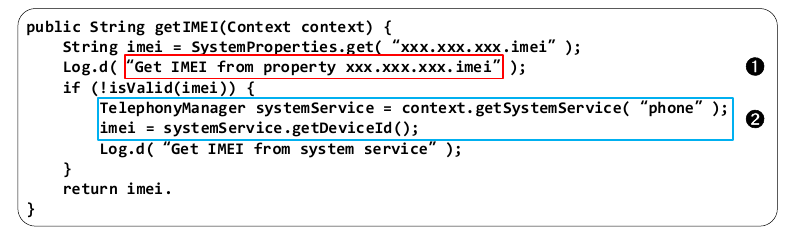}
  }
  \caption{Key Information for Identifying Property Content.}
  \label{fig:manuel}
\end{figure}

In this way, we narrow down the range of system properties and settings that might contain sensitive identifiers.
Subsequently, we adopt manual inspection to finally confirm which system properties/settings contain non-resettable device identifiers leveraging the context code information.
In these code, there is a lot of information that could helps us determine the content of system properties or settings. 
For example, some log information contains descriptions of code behavior, as shown in the~\autoref{fig:manuel} \ding{202}.
Additionally, there may be other pieces of code with the same/similar functionality, as shown in~\autoref{fig:manuel} \ding{203}.
The code in~\autoref{fig:manuel} first tries to obtain the content from the system property. 
If the content is not valid, it then attempts to retrieve the device's IMEI from the system service. 
By understanding the semantics, we can infer that the system property's content should be consistent with the content retrieved from the system service, thereby determining that the system property stores the device's IMEI.
Our manual inspection process is primarily led by two experienced Android researchers, with a third for dispute resolution. 
First, we have two researchers analyze the same context information and determine which system properties and settings store sensitive information.
When the two researchers' opinions coincide, we directly adopt the result, and whenever there are different opinions, the two researchers discuss with the third researcher until a consensus is reached.

It is worth noting that during our manual analysis, we identify two situations that could potentially lead to inaccurate results.
(1) We find that some system properties/settings are accessed by only third-party pre-installed apps, which try to obtain non-resettable device identifiers through various methods, with system properties or settings being just one of them, like the example shown in~\autoref{fig:manuel}. 
We speculate that these third-party apps are uncertain whether the specific system property or setting exist in custom systems and are merely attempting to retrieve device identifiers from them. 
To reduce inaccuracies arising from such situations, we use a method similar to our empirical study by using the ``strings'' command to extract strings within all kinds of files (e.g., SO libraries and executable files) in the same custom ROM, and check for the presence of any that contain non-resettable device identifiers we identified. 
Through this method, we believe that only system properties or settings that can be identified in system components, determined by file paths, are relatively accurate.
(2) Some third-party SDKs in system apps access system properties that obviously do not exist in the system. 
For instance, we find access to the ``xxx.xxx.miui.xxx'' system property in system apps within the Meizu custom system, where MIUI is the name of Xiaomi's custom system. 
We manually exclude these cases for accuracy.

\subsection{Access Control Analysis}
\label{sec:approachaccess}
Finally, we analyze the access control policies for system properties and settings storing non-resettable device identifiers to find those lacking effective access control.
As mentioned in~\S\ref{sec:background2}, access control for system properties is primarily enforced through SELinux. 
We begin by identifying the property context of each custom system property, which is stored in files whose names end with ``property\_contexts''. We directly search for these files in custom ROMs.
Next, we examine the contents of these files to determine the appropriate property contexts for the system properties that hold sensitive device identifiers, with a particular focus on the ``type'' domain, such as ``system\_id\_prop'' shown in~\autoref{tab:selinux}.
Note that in the property context, property name can include the wildcard character ``*'', e.g., ``ro.*'' represents properties starting with ``ro.''.
The use of wildcards may result in a property matching multiple property contexts. 
For instance, ``ro.xxx.xxx.imei1'' could align with both the specific context ``ro.xxx.xxx.imei1'' and the more general pattern ``ro.*''. 
In such instances, SELinux prioritizes the context that is the most specific match. Typically, the property context name with the longest match to the property name is selected as the best fit.

After identifying the property context of a system property, we analyze the associated type attribute sets and policy rules in policy binary files (pre-Android 8.0) or CIL files (post-Android 8.0). 
CIL files are text files whose contents can be directly read, whereas, for policy binary files, we use the SETools~\cite{setools} to parse the embedded rules.
We locate all policy files and CIL files in custom ROMs, and then iterate through SELinux rules twice.
In the first iteration, we record all details related to type attribute sets, including definitions and the use of the ``expandtypeattribute'' keyword.
Given that the definition of a type attribute set may appear after its usage, and a set might reference other sets across various CIL files, we postpone specific policy rule analysis during the first pass.
Instead, we focus on identifying the types encompassed within each set after a comprehensive review of all CIL files.
More specifically, we parse the definitions of type attribute sets to determine the included and excluded elements (i.e., types or sets). 
Upon completing the file review, we assess each element. 
Specific types or non-expandable sets are immediately categorized into included or excluded groups.
For expandable sets, we further dissect the types they encompass, ultimately incorporating these into the corresponding group. 
Finally, we determine the types in each set by taking all types that are in the included set but not in the excluded set.

During the second iteration, we examine all policy rules, focusing on the source and target in each rule. 
If the target of a rule is the type to which a sensitive system property belongs, as determined by the analysis of property context, or a type attribute set encompassing this type, we consider the rule applicable to the sensitive property.
We also analyze the source in the rules to identify which types of entities, such as ``system\_app'' in~\autoref{tab:selinux}, have access these sensitive properties.
Generally, third-party apps belong to the ``untrusted\_app'' type. 
If, through analyzing SELinux policies, we find that a system property can be accessed by the ``untrusted\_app'' type, we conclude that third-party apps have access to it.

On the other hand, as described in~\S\ref{sec:background2}, to analyze the access control policies for reading system settings, we focus on the definitions and annotations of system settings in the \texttt{\seqsplit{android.provider.Settings}} class, typically located in ``framework.jar'' file. 
Using Soot, we analyze all fields in the ``Settings.Global'', ``Settings.System'', and ``Settings.Secure'' inner classes, corresponding to the three categories of Settings.
We extract the field names and their annotations, such as ``@Readable'' and ``@SystemApi'', which are our primary focus.
This process helps us determine the access control policies for system settings. 
Specifically, settings not explicitly defined in the Settings class are accessible by all system entities, including third-party apps.
Prior to Android 12, any system setting without the ``@SystemApi'' annotation is accessible to third-party apps. 
From Android 12 onwards, only those settings marked ``@Readable'' and not ``@SystemApi'' are accessible by third-party apps.

%% file: Sections/results.tex
\section{Results}
\label{sec:result}

\subsection{Dataset}

\begin{table*}[th]
\caption{The Distribution of Our Custom System Dataset and Summary of Our Results.}
\label{tab:dataset}
\begin{threeparttable}
\resizebox{1\linewidth}{!}{
\begin{tabular}{|c|c|ccccccccc|cc|cc|cc|}
\hline
\multirow{2}{*}{Brand} & \multirow{2}{*}{\# Devices} & \multicolumn{9}{c|}{Android Version}                   & \# Sensitive & \# (\%) Vulnerable & \# Sensitive & \# (\%) Vulnerable & \# (\%) Sensitive & \# (\%) Vulnerable \\ 
                       &                             & Pre-v6 & v7  & v8  & v9  & v10 & v11 & v12 & v13 & v14 & Properties   & Properties         & Settings     & Settings           & Devices           & Devices            \\ \hline \hline
Alps                   & 26                            & \multicolumn{1}{c|}{3}      & \multicolumn{1}{c|}{3}   & \multicolumn{1}{c|}{7}   & \multicolumn{1}{c|}{4}   & \multicolumn{1}{c|}{5}   & \multicolumn{1}{c|}{0}   & \multicolumn{1}{c|}{3}   & \multicolumn{1}{c|}{1}   & 0   & 94           & 40 (43\%)          & 24           & 5 (21\%)           & 10 (38\%)         & 7 (27\%)           \\ \hline
Asus                   & 55                            & \multicolumn{1}{c|}{1}      & \multicolumn{1}{c|}{6}   & \multicolumn{1}{c|}{10}  & \multicolumn{1}{c|}{11}  & \multicolumn{1}{c|}{11}  & \multicolumn{1}{c|}{3}   & \multicolumn{1}{c|}{3}   & \multicolumn{1}{c|}{10}  & 0   & 36           & 22 (61\%)          & 35           & 3 (9\%)            & 12 (22\%)         & 6 (11\%)           \\ \hline
Digma                  & 25                            & \multicolumn{1}{c|}{0}      & \multicolumn{1}{c|}{1}   & \multicolumn{1}{c|}{20}  & \multicolumn{1}{c|}{4}   & \multicolumn{1}{c|}{0}   & \multicolumn{1}{c|}{0}   & \multicolumn{1}{c|}{0}   & \multicolumn{1}{c|}{0}   & 0   & 20           & 3 (15\%)           & 35           & 0 (0\%)            & 25 (100\%)        & 1 (4\%)            \\ \hline
Huawei                 & 29                            & \multicolumn{1}{c|}{2}      & \multicolumn{1}{c|}{1}   & \multicolumn{1}{c|}{0}   & \multicolumn{1}{c|}{5}   & \multicolumn{1}{c|}{7}   & \multicolumn{1}{c|}{5}   & \multicolumn{1}{c|}{9}   & \multicolumn{1}{c|}{0}   & 0   & 6            & 1 (17\%)           & 5            & 1 (20\%)           & 5 (17\%)          & 1 (3\%)            \\ \hline
Lenovo                 & 99                            & \multicolumn{1}{c|}{21}     & \multicolumn{1}{c|}{9}   & \multicolumn{1}{c|}{8}   & \multicolumn{1}{c|}{26}  & \multicolumn{1}{c|}{12}  & \multicolumn{1}{c|}{13}  & \multicolumn{1}{c|}{6}   & \multicolumn{1}{c|}{4}   & 0   & 529          & 337 (64\%)         & 182          & 51 (28\%)          & 85 (86\%)         & 75 (76\%)          \\ \hline
Meizu                  & 40                            & \multicolumn{1}{c|}{10}     & \multicolumn{1}{c|}{6}   & \multicolumn{1}{c|}{8}   & \multicolumn{1}{c|}{6}   & \multicolumn{1}{c|}{2}   & \multicolumn{1}{c|}{4}   & \multicolumn{1}{c|}{0}   & \multicolumn{1}{c|}{3}   & 1   & 218          & 125 (56\%)         & 142          & 44 (31\%)          & 39 (97\%)         & 37 (93\%)          \\ \hline
Motorola               & 144                           & \multicolumn{1}{c|}{9}      & \multicolumn{1}{c|}{11}  & \multicolumn{1}{c|}{17}  & \multicolumn{1}{c|}{24}  & \multicolumn{1}{c|}{21}  & \multicolumn{1}{c|}{13}  & \multicolumn{1}{c|}{33}  & \multicolumn{1}{c|}{16}  & 0   & 416          & 162 (39\%)         & 146          & 55 (38\%)          & 118 (82\%)        & 113 (78\%)         \\ \hline
Nokia                  & 88                            & \multicolumn{1}{c|}{4}      & \multicolumn{1}{c|}{1}   & \multicolumn{1}{c|}{5}   & \multicolumn{1}{c|}{17}  & \multicolumn{1}{c|}{30}  & \multicolumn{1}{c|}{10}  & \multicolumn{1}{c|}{9}   & \multicolumn{1}{c|}{12}  & 0   & 363          & 266 (73\%)         & 100          & 36 (36\%)          & 49 (56\%)         & 40 (45\%)          \\ \hline
Nubia                  & 29                            & \multicolumn{1}{c|}{0}      & \multicolumn{1}{c|}{3}   & \multicolumn{1}{c|}{3}   & \multicolumn{1}{c|}{8}   & \multicolumn{1}{c|}{3}   & \multicolumn{1}{c|}{5}   & \multicolumn{1}{c|}{4}   & \multicolumn{1}{c|}{3}   & 0   & 78           & 46 (59\%)          & 87           & 25 (28\%)          & 29 (100\%)        & 21 (72\%)          \\ \hline
Oneplus                & 45                            & \multicolumn{1}{c|}{2}      & \multicolumn{1}{c|}{0}   & \multicolumn{1}{c|}{0}   & \multicolumn{1}{c|}{3}   & \multicolumn{1}{c|}{8}   & \multicolumn{1}{c|}{17}  & \multicolumn{1}{c|}{8}   & \multicolumn{1}{c|}{7}   & 0   & 195          & 106 (54\%)         & 78           & 47 (60\%)          & 38 (84\%)         & 36 (80\%)          \\ \hline
Oppo                   & 64                            & \multicolumn{1}{c|}{6}      & \multicolumn{1}{c|}{2}   & \multicolumn{1}{c|}{8}   & \multicolumn{1}{c|}{9}   & \multicolumn{1}{c|}{10}  & \multicolumn{1}{c|}{5}   & \multicolumn{1}{c|}{6}   & \multicolumn{1}{c|}{18}  & 0   & 384          & 229 (60\%)         & 145          & 25 (17\%)          & 61 (95\%)         & 59 (92\%)          \\ \hline
Poco                   & 22                            & \multicolumn{1}{c|}{0}      & \multicolumn{1}{c|}{0}   & \multicolumn{1}{c|}{0}   & \multicolumn{1}{c|}{0}   & \multicolumn{1}{c|}{3}   & \multicolumn{1}{c|}{8}   & \multicolumn{1}{c|}{6}   & \multicolumn{1}{c|}{5}   & 0   & 363          & 69 (19\%)          & 120          & 69 (56\%)          & 22 (100\%)        & 22 (100\%)         \\ \hline
Qti                    & 42                            & \multicolumn{1}{c|}{0}      & \multicolumn{1}{c|}{0}   & \multicolumn{1}{c|}{0}   & \multicolumn{1}{c|}{2}   & \multicolumn{1}{c|}{2}   & \multicolumn{1}{c|}{6}   & \multicolumn{1}{c|}{28}  & \multicolumn{1}{c|}{4}   & 0   & 326          & 132 (40\%)         & 173          & 93 (54\%)          & 41 (98\%)         & 40 (95\%)          \\ \hline
Realme                 & 65                            & \multicolumn{1}{c|}{0}      & \multicolumn{1}{c|}{0}   & \multicolumn{1}{c|}{1}   & \multicolumn{1}{c|}{8}   & \multicolumn{1}{c|}{19}  & \multicolumn{1}{c|}{11}  & \multicolumn{1}{c|}{10}  & \multicolumn{1}{c|}{15}  & 1   & 396          & 192 (48\%)         & 137          & 37 (27\%)          & 65 (100\%)        & 65 (100\%)         \\ \hline
Redmi                  & 59                            & \multicolumn{1}{c|}{0}      & \multicolumn{1}{c|}{0}   & \multicolumn{1}{c|}{0}   & \multicolumn{1}{c|}{0}   & \multicolumn{1}{c|}{7}   & \multicolumn{1}{c|}{10}  & \multicolumn{1}{c|}{11}  & \multicolumn{1}{c|}{26}  & 5   & 673          & 124 (18\%)         & 208          & 100 (48\%)         & 59 (100\%)        & 54 (92\%)          \\ \hline
Samsung                & 158                           & \multicolumn{1}{c|}{20}     & \multicolumn{1}{c|}{11}  & \multicolumn{1}{c|}{13}  & \multicolumn{1}{c|}{28}  & \multicolumn{1}{c|}{27}  & \multicolumn{1}{c|}{25}  & \multicolumn{1}{c|}{15}  & \multicolumn{1}{c|}{19}  & 0   & 534          & 226 (42\%)         & 524          & 212 (40\%)         & 154 (97\%)        & 134 (85\%)         \\ \hline
Xiaomi                 & 146                           & \multicolumn{1}{c|}{9}      & \multicolumn{1}{c|}{12}  & \multicolumn{1}{c|}{24}  & \multicolumn{1}{c|}{11}  & \multicolumn{1}{c|}{25}  & \multicolumn{1}{c|}{26}  & \multicolumn{1}{c|}{4}   & \multicolumn{1}{c|}{29}  & 6   & 1598         & 441 (28\%)         & 647          & 281 (43\%)         & 137 (94\%)        & 116 (79\%)         \\ \hline
Zte                    & 28                            & \multicolumn{1}{c|}{2}      & \multicolumn{1}{c|}{2}   & \multicolumn{1}{c|}{3}   & \multicolumn{1}{c|}{11}  & \multicolumn{1}{c|}{5}   & \multicolumn{1}{c|}{3}   & \multicolumn{1}{c|}{2}   & \multicolumn{1}{c|}{0}   & 0   & 153          & 116 (76\%)         & 188          & 57 (30\%)          & 23 (82\%)         & 15 (54\%)          \\ \hline
Other (232)            & 650                           & \multicolumn{1}{c|}{71}     & \multicolumn{1}{c|}{42}  & \multicolumn{1}{c|}{198} & \multicolumn{1}{c|}{148} & \multicolumn{1}{c|}{80}  & \multicolumn{1}{c|}{65}  & \multicolumn{1}{c|}{30}  & \multicolumn{1}{c|}{14}  & 2   & 1810         & 840 (46\%)         & 644          & 195 (30\%)         & 434 (67\%)        & 270 (42\%)         \\ \hline
Total (250)            & 1814                          & \multicolumn{1}{c|}{160}    & \multicolumn{1}{c|}{110} & \multicolumn{1}{c|}{325} & \multicolumn{1}{c|}{325} & \multicolumn{1}{c|}{277} & \multicolumn{1}{c|}{229} & \multicolumn{1}{c|}{187} & \multicolumn{1}{c|}{186} & 15  & 8192         & 3477 (42\%)        & 3620         & 1336 (37\%)        & 1406 (78\%)       & 1112 (61\%)        \\ \hline
\end{tabular}
}
\begin{tablenotes}
	\footnotesize
	\item Note: we separately summarize the 18 brands with more than 20 devices each, while the remaining 232 brands were combined in the statistics. In addition, due to the small \\number of ROMs before Android 6, we combine the statistics for ROMs from Android 6 and earlier versions. We also combine the sub-versions, such as Android 8.0 and 8.1. \\The percentages in the Vulnerable Properties and Settings columns are calculated by \# Vulnerable/\# Sensitive, and the percentages in the Sensitive/Vulnerable Devices \\columns are calculated in comparison to the \# Devices.
\end{tablenotes}
\end{threeparttable}
\end{table*}

To facilitate our large-scale investigation, we collected custom Android ROMs from the public project Android Dumps~\cite{android_dumps}, which hosts a wide range of Android stock ROMs.
In October 2023, we collected all system ROMs from this source, resulting in a dataset of \datasetsize ROMs from \totalbrand vendors.
The vendors, device numbers, and Android API versions of these ROMs are summarised in~\autoref{tab:dataset}.
Our dataset primarily includes ROMs from major Android vendors such as Samsung, Xiaomi, Huawei, Lenovo, 
which constitute the majority of the dataset.
It also includes ROMs from smaller vendors like Gionee and Blackview, usually with only 2-3 ROMs each.
Notably, our dataset includes Google’s ROMs, such as those for Pixel devices, which feature Google's customizations. 
The ROMs in the dataset span Android versions from 4.0.3 to 14, with build dates ranging from 2013 to 2023.
This diverse dataset enables us to thoroughly explore the covert channels used to access non-resettable device identifiers in custom Android systems.
Note that the ROMs from Android Dumps have already been parsed into specific file collections rather than packaged as image files, allowing us to directly analyze the files without additional unpacking work. 

\subsection{Custom System Properties/Settings with Non-Resettable Identifiers}
\label{sec:staticresult}

Applying {\framework} to all custom ROMs we collected, we successfully analyzed over 600K APK and JAR files within these custom systems, generating over 270GB of result data. 
This data includes accessed system properties and settings, as well as extensive contextual code information.
After applying the filtering process described in~\S\ref{sec:approachfilter}, we identified about 30K potential cases.
Over the course of five days, two experienced Android researchers completed the manual verification on these cases. 
During this process, we observed that different ROMs from the same brand often shared repeated cases, including identical system property or setting names and contextual code.
We believe that these properties and settings are the same across ROMs of the same brand. 
Consequently, we avoid duplicate manual reviews of these cases, which significantly reduces our workload.
As a result, we confirmed a total of \totalsensitivepropertiescase system properties and \totalsensitivesettingscase system settings containing non-resettable device identifiers, including \totalsensitiveproperties unique system properties and \totalsensitivesettings system settings\footnote{The same system property or setting may recur across different ROMs, resulting in a total count significantly higher than the number of unique types.}.
We found at least one instance of these in \totalsensitivedevices custom ROMs. 
The detailed results are presented in~\autoref{tab:dataset}, with a breakdown by identifier type available in~\autoref{tab:typesplit} in appendices.
For clarity, we will refer to system properties and settings that store non-resettable device identifiers as \textbf{sensitive system properties and settings} in the remainder of the text.
Our results reveal that sensitive system properties and settings introduced through system customizations are widely prevalent.

\begin{figure}[h]
  \centering
  \resizebox{1\linewidth}{!}{
  \includegraphics[width=\textwidth]{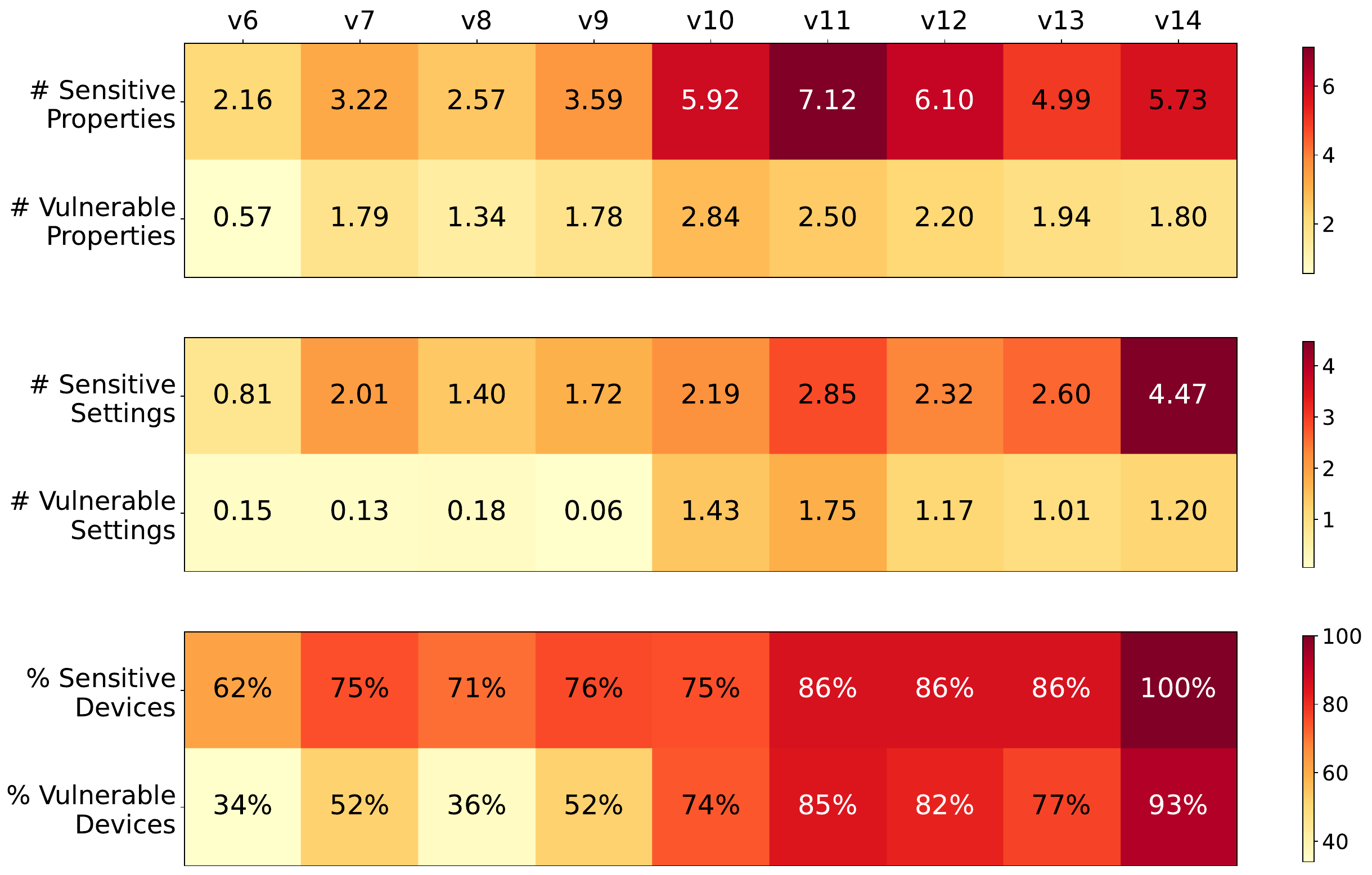}
  }
  \caption{The Average of System Properties/Settings and  Percentage of Devices across Different System Versions.}
  \label{fig:staticversion}
\end{figure}

We also examined the evolution of sensitive system properties/settings across different Android versions.
As shown in~\autoref{fig:staticversion}, from the early Android versions up to Android 10, the average of sensitive system properties and settings in custom systems gradually increased, with the most significant growth occurring in Android 10. 
We speculate that this increase is related to Android's tightening policies on device identifiers, particularly the complete prohibition of third-party apps from using non-resettable device identifiers in Android 10.
After Android 10, there was a slight decline, possibly influenced by heightened regulatory scrutiny from both societal and academic spheres.
It is worth noting that our dataset includes only 15 custom ROMs for Android 14, so the small sample size may have skewed the data for this version, making it an outlier compared to the overall trend.

Additionally, we investigate which supply chain actors utilize covert channels to access identifiers and their purposes for doing so. 
We identified these actors by analyzing package names associated with the code accessing sensitive system properties and settings. We then focused on frequently occurring package names for a detailed case study, manually examining their code.
Our findings reveal that custom system developers (e.g., Oppo~\cite{oppo}) and hardware providers (e.g., MediaTek~\cite{mediatek}) use these properties within system components to monitor device status. 
For example, MediaTek's telephony component checks the IMSI or ICCID stored in system properties to determine if a SIM card is present.
Additionally, various SDKs in system apps leverage these properties and settings. 
This includes SDKs from device vendors like Xiaomi's~\cite{xiaomi} analytics and crash reporting SDKs, Lenovo's~\cite{lenovo} push SDK, and third-party SDKs such as Baidu's~\cite{baidu} map SDK and Alibaba's~\cite{alibaba} security SDK.
They use these identifiers for device identification, transmitting user usage data, crash reports, and identifier to remote endpoints.

Furthermore, our analysis shows that in higher versions of Android, supply chain actors increasingly prefer using custom system properties or settings instead of official APIs to access device identifiers. 
This trend reinforces our hypothesis that, despite having system-level permissions, these actors are affected by system modifications that disrupt stable user tracking via official APIs. 
As a result, they turn to covert channels to ensure consistent tracking.

\noindent \textbf{Our Findings.}
\textit{A substantial number of system properties and settings storing non-resettable device identifiers are present in custom systems across various brands and system versions. 
The introduction of these properties/settings is likely driven by the need for convenient access to identifiers, in response to restrictions on their usage.}

\subsection{Vulnerable Implementations}
Next, we analyzed the access control policies for these sensitive system properties and system settings following the method described in~\S\ref{sec:approachaccess} to see if they have proper access control.
For the \totalsensitivepropertiescase system properties and \totalsensitivesettingscase system settings, we identified \totaldangerouspropertiescase system properties and \totaldangeroussettingscase system settings without proper access control, allowing any third-party apps to access these identifiers. We found at least one such system property or setting in \totaldangerousdevices custom ROMs.
Similarly, for the sake of simplicity, we refer to sensitive system properties and settings lacking effective access control as \textbf{vulnerable system properties and settings} in the following.
Our results indicate that vulnerable system properties and settings are prevalent in custom systems.
It is worth mentioning that we did not find any such vulnerabilities in Google's systems, demonstrating Google's robust code security that aligns with findings from previous studies~\cite{elsabagh2020firmscope, hou2022large}.
In terms of the trend across system versions, it is also quite similar to that of sensitive system properties and settings, as shown in~\autoref{fig:staticversion}.
Before Android 10, due to the relatively small number of sensitive system properties and settings, the difficulty of implementing access control was lower, resulting in more effective access control.
After Android 10, as the number of sensitive system properties and settings increased, the implementation of access control policies became more prone to errors.
Furthermore, we analyzed some of these cases with vulnerable access control, and summarized some of the possible reasons that may lead to vulnerable implementations.

\noindent \textbf{Overlooked System Properties and Settings.}
We believe that the most direct cause of vulnerable access control is the oversight during development, where appropriate access control policies were not deployed for sensitive system properties and settings.
For example, in the custom ROM of BRAND-A (Android version 14), we have reported four vulnerable system properties ``xx.xx.xx.imei1'', ``xx.xx.xx.imei2'', ``xx.xx.xx.meid'', and ``xx.xx.xx.sn'', which have been confirmed by the vendors.
Our analysis of all property contexts in this system revealed that the most common matching context name for these properties is ``*'', the default property context type. 
This rule can match any system property and also allows any third-party apps to access such system properties. 
In this example, we consider that the developers forgot to add a property context for this sensitive system property, resulting in the vulnerable access control.
There are a total of 844 similar cases in our results where the most matching property context for a sensitive system property is the default context (i.e., the property name in the security context is ``*''). 
On the other hand, most of the vulnerable system settings (i.e., 95\%) we discovered are not defined (overlooked) in the Settings class. 
This lack of definition allows third-party apps to access these sensitive settings, leading to access control issues.

Another reason sensitive system properties and settings may be overlooked is the large number of supply chain actors involved in custom systems.
When establishing access control policies, developers may focus on the custom properties and settings introduced by major supply chain actors, including device vendors and hardware providers.
However, they might miss some covert third-party actors, like software vendors, who also introduce sensitive properties and settings, leading to vulnerabilities.
We use the two sensitive system settings, ``xxx\_xxx\_i'' and ``xxx.xxx.deviceid'', introduced by Baidu~\cite{baidu}, as an example.
The ``xxx\_xxx\_i'' setting stores the device's plaintext IMEI, while ``xxx.xxx.deviceid'' contains an encrypted key-value pair that includes the IMEI and another identifier generated using the android\_id~\cite{androidid}. 
The encryption uses AES in CBC mode, but with a key hard-coded in the code, and the final string is encoded in base64.
These two settings are widely distributed across custom systems.
Our analysis revealed that access to these sensitive settings primarily occurs within the Baidu SDK used in many system apps, such as app stores, map apps, browsers, and weather apps.
In our results, we identified a total of 3,120 apps that access these settings. 
The widespread presence of these sensitive settings in custom systems, combined with the lack of access control in all the systems we analyzed, has significantly impacted our results (as shown in~\autoref{tab:dataset} and~\autoref{fig:staticversion}).

\noindent \textbf{Overly Complex Access Control Rules.}
SELinux policies in custom systems may contain a large number of rules due to the extensive customizations.
For example, in the custom ROM of BRAND-B (Android version 14), there are over 2,500 property contexts and more than 50,000 lines of specific SELinux policy files.
We have reported two vulnerable system properties ``xxx.xxx.xxx.btmac'' and ``xxx.xxx.xxx.wifimac'' identified in this ROM, which was also confirmed by device vendors.
These properties belong to ``radio\_prop'' type, which is included in six type attribute sets and linked to dozens of policy rules.
One of these rules allows ``untrusted\_app'' to access it, resulting in the access control issue for this sensitive property.
Therefore, we believe that overly complex access control rules contribute to such vulnerabilities.
On the one hand, as in the case of the above example, involving too many type attribute sets can increase the probability of introducing access control vulnerabilities. 
This is because a mistake in any of these sets can lead to access control errors for the system property. 
On the other hand, an excessive number of rules degrade the auditability of access control, making it more difficult to identify existing access control issues.

\noindent \textbf{Additional Access Channels for System Properties.}
Generally, the methods for accessing system properties or settings are through system APIs or commands, as mentioned in~\S\ref{sec:background1}.
However, in our findings, we discovered that some sensitive system properties are accessed by system services and returned by methods within these system services, thereby introducing additional access channels for sensitive system properties.
To investigate such issues, we conducted additional analysis on the access behaviors of sensitive system properties within system services. 
Since only specific system files are loaded as system services~\cite{systemservice}, we can identify potential access behaviors of sensitive system properties within system services by examining the file paths where these behaviors occur.
After identifying the potential targets, we further confirm whether the behavior occurs within a system service by examining the parent class of the class where the sensitive system property behavior is located. 
Typically, classes that host system services typically extend the \texttt{\seqsplit{com.android.server.SystemService}} class or the Binder's Stub class~\cite{binder}  (any class using Binder as a receiver also subclasses this class).
Finally, we manually analyzed the obtained results to: 
(1) confirm whether the class containing the sensitive system property access behavior was part of a system service; 
(2) determine whether the sensitive system property was accessed within a method of the system service and returned as a method result; 
(3) investigate whether the method was a public method and whether there were any access control policies in place to prevent access by third-party apps.
As a result, we found eight related cases.

\begin{figure}[h]
  \centering
  \resizebox{1\linewidth}{!}{
  \includegraphics[width=\linewidth]{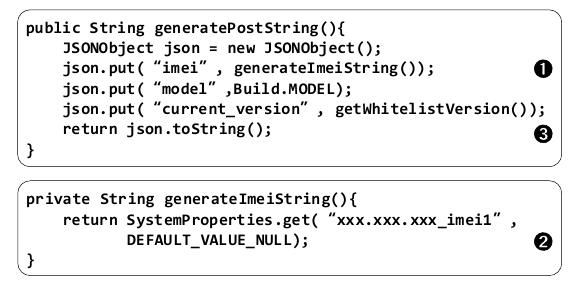}
  }
  \caption{The Code of the Vulnerable System Service.}
  \label{fig:service}
\end{figure}

An example is the system service named ``xxxService'' in the custom ROM of BRAND-C (Android version 12).
Within this service, there is a method with the signature ``public String generatePostString()''. The code for this method and its related methods is shown in~\autoref{fig:service}.
The method ``generatePostString'' further calls a method named ``generateImeiString'', as shown in Step~\ding{202}. 
This method reads the value of the sensitive but not vulnerable system property named ``xxx.xxx.xxx\_imei1'', as illustrated in Step~\ding{203}. 
The method ``generatePostString'' retrieves the device's IMEI, puts it into a JSONObject, and then returns the content of sensitive system property with other device information, as shown in Step~\ding{204}.
Since ``xxxService'' is a system service, and the ``generatePostString'' method is public with no additional access control, untrusted third-party apps can access this method and retrieve the device's IMEI.
One exploitation method for this type of case is to access the system service directly via Java reflection and Binder mechanism, which does not require the application context.
The additional channels for accessing sensitive system properties introduced in system services without access control result in the vulnerabilities.

\noindent \textbf{Our Findings.}
\textit{Vulnerable sensitive system properties and settings are prevalent in custom Android systems, and the reasons can be attributed to extensive and complex system customizations, as well as developer oversights.}

\subsection{Result Verification}
We further verify the accuracy of the vulnerable system properties and settings we identified, by generating Proof of Concepts (PoCs) on real smartphones.
Due to the limited availability of testing devices from various manufacturers and devices, we chose to take advantage of remote real-device testing services provided by device vendors and commercial platform (i.e., Alibaba Cloud~\cite{aliyun}). 
These services provide access to a wide range of models through remote desktop access, allowing users to upload and run test apps.
However, in our dataset, there is only one version of the custom ROM for each device, corresponding to a specific system version. 
The remote testing services might not have the exact brand and device we need, and even if they do, the system version might differ from our dataset.
Therefore, within the practical constraints, we are unable to generate PoCs for all identified vulnerabilities. 
As an alternative, we found 32 devices in the remote real-device testing service that have the same device model and Android system version as those in our dataset (although the specific build versions differ, such as the custom ROM compiled in September 2023 in our dataset versus one compiled in October 2023 in the testing service). 
We used these devices to evaluate the accuracy of our detection results.
Specifically, we developed a testing app without any permission requests that functions as a third-party app to retrieve all system properties and settings on the device, and check for the presence of any vulnerable system properties or settings.
We manually compared the contents of the found properties and settings with the non-resettable device identifiers provided in the devices' settings app, confirming that they indeed contained non-resettable identifiers.

In summary, our approach identified 130 vulnerable properties and settings across the 32 devices, and we successfully generated PoCs for 102 of them. 
After verification, all 102 properties and settings were confirmed to store non-resettable device identifiers.
For the remaining 28 cases, our manual verification shows that they are primarily caused by one system settings, ``xxx\_device\_mac''.
The cases related to ``xxx\_device\_mac'' were caused by the MediaTek~\cite{mtk} telephony service component. 
Upon analyzing its code, we found that it attempts to obtain the device MAC address from various sources, with accessing ``xxx\_device\_mac'' being one of the methods.
These cases can only be triggered under specific configurations, thus we cannot generate PoCs for them automatically.
We will discuss these limitations in~\S\ref{sec:limitations}.
Overall, the on-device dynamic evaluation demonstrates the high accuracy of our approach.

\noindent \textbf{Our Findings.}
\textit{Dynamic exploration of real-world devices shows the great accuracy of our approach. We can generate PoCs for most of the vulnerable system properties or settings automatically, while the remaining ones can only be triggered under specific conditions.
}

\subsection{Comparison with State-of-the-art}
We further compare our approach with U2-I2~\cite{meng2023post}, the most closely related work.
In terms of research scale, U2-I2 covered only 13 devices, identifying 30 vulnerable system properties and settings. 
In contrast, our dataset and the total number of identified issues are two orders of magnitude larger, allowing for a comprehensive examination of non-resettable identifiers in system properties and settings, with broader coverage in both time scale and device diversity.
Besides, we further compare the capability of these two approaches for analyzing the same ROM.
Upon reviewing its open-source artifact~\cite{gdprartifact}, we find that the test devices have been anonymized, making it impossible for us to conduct a comparison against their results.
Additionally, we are unable to apply the U2-I2 method on remote real-devices, as factory resets, which are necessary for the U2-I2 method, cannot be performed on these devices.
Ultimately, we have gathered four real devices that can be used for fair comparison. For these devices, their ROMs are included in our database.
Further, we conducted dynamic testing on these devices using the U2-I2 tool.
Specifically, following the guidelines, we searched for non-resettable identifiers by running the U2-I2 tester app before and after factory reset of devices and comparing the results.
We ran the ``GETID'' functionality in the tool, which is relevant to our research objectives and is used to detect non-resettable identifiers in system properties and settings.
During the testing, we found that the process required considerable manual effort. 
Since a factory reset was involved, we first had to restore rooted test device to prevent it from becoming bricked (i.e., can no longer start normally).
Additionally, parts of the tool’s testing process, including the factory reset, re-enabling of ADB debugging post-reset, determination of system properties/settings, and final result comparison, could not be automated.
In particular, during the result comparison, many items remained unchanged after the reset, even after excluding the very short ones. 
It is necessary to check the device's non-resettable identifiers in the settings app and compare them with the unchanged system properties and settings.

As a result, U2-I2 identified three non-resettable identifiers in the system properties and settings across four devices, while our method not only detected these three cases but also found two more identifiers in system settings on two devices, namely ``xxx\_sim\_imsi'' and ``xxx\_sim\_iccid''.
These two system settings are associated with a feature related to mobile data and only appear after activating specific functions on the device. 
Due to the lack of complete feature setup in the U2-I2 method, it failed to detect them.
This highlights U2-I2's limitation in detecting issues that rely on specific preconditions (e.g., activating a feature or performing an action), whereas our static approach is capable of uncovering such vulnerabilities.
However, we identified one system setting (the aforementioned ``xxx\_device\_mac'') through our method that was not found on the device, indicating a potential false positive.
In summary, our method not only replicates U2-I2’s findings with acceptable accuracy loss, but also uncovers additional vulnerabilities.

\noindent \textbf{Our Findings.}
\textit{
Compared to existing work, our experiments operate on a much larger scale without any limitation, enabling a thorough examination of non-resettable identifiers across diverse devices.
Even testing on the same devices, our method can uncover more vulnerabilities that were not detected by U2-I2. It suggests the high scalability and accuracy of our approach.
}

\subsection{Brand-Wide Recurring Vulnerabilities}
As mentioned in~\ref{sec:staticresult}, we observed that the same vulnerabilities often appeared on multiple devices from the same brand.
Inspired by this, we extended our testing to \dynamicmodels new devices of the brands we have analyzed, using the remote real-device service.
Through this method, we expect to identify more vulnerable devices, even their system ROMs are not available.

\begin{table}[t]
\caption{The Device Information of Extended Evaluation.}

\label{tab:dynamic}
\begin{threeparttable}
\resizebox{1\linewidth}{!}{
\begin{tabular}{|c|c|cccccccccc|c|c|}
\hline
\multirow{2}{*}{Brand} & \multirow{2}{*}{\# Devices} & \multicolumn{10}{c|}{Android Version}                 & \multirow{2}{*}{\# VP} & \multirow{2}{*}{\# VS} \\
                       &                             & v6 & v7 & v8 & v9 & v10 & v11 & v12 & v13 & v14 & v15 &                        &                        \\ \hline
Huawei                 & 29/11                       & 0  & 0  & 3  & 3  & 20  & 0   & 3   & 0   & 0   & 0   & 0                      & 22                     \\ \hline
Motorola               & 7/7                         & 1  & 3  & 0  & 0  & 0   & 3   & 0   & 0   & 0   & 0   & 18                     & 11                     \\ \hline
Oppo                   & 53/31                       & 0  & 0  & 0  & 2  & 5   & 10  & 6   & 17  & 13  & 0   & 4                      & 56                     \\ \hline
Oneplus                & 11/7                        & 0  & 0  & 0  & 0  & 0   & 0   & 0   & 0   & 9   & 2   & 0                      & 15                     \\ \hline
Samsung                & 35/21                       & 0  & 2  & 2  & 3  & 3   & 0   & 10  & 9   & 6   & 0   & 12                     & 50                     \\ \hline
Vivo                   & 110/97                      & 0  & 0  & 4  & 2  & 7   & 24  & 18  & 28  & 26  & 1   & 36                     & 183                    \\ \hline
Xiaomi                 & 20/9                        & 0  & 0  & 0  & 0  & 0   & 0   & 0   & 12  & 8   & 0   & 4                      & 9                      \\ \hline
Redmi                  & 23/13                       & 0  & 0  & 0  & 0  & 0   & 1   & 2   & 17  & 3   & 0   & 10                     & 12                     \\ \hline
Other (14)             & 26/20                       & 1  & 6  & 3  & 4  & 4   & 4   & 3   & 0   & 0   & 1   & 76                     & 42                     \\ \hline
Total (22)             & 314/216                     & 2  & 11 & 12 & 14 & 39  & 42  & 42  & 83  & 65  & 4   & 160                    & 400                    \\ \hline
\end{tabular}
}
\begin{tablenotes}
	\footnotesize
	\item Note: we separately summarize the 8 brands with more than 5 devices each, whi-\\le the remaining 14 brands were combined in the statistics. The ``Devices'' column \\contains two numbers: the total number of devices tested and the number of \\devices with vulnerable system properties or settings. The last two columns, ``VP'' \\and ``VS'', represent vulnerable properties and vulnerable settings, respectively.
\end{tablenotes}
\end{threeparttable}
\end{table}

Our dynamic testing results, summarized in~\autoref{tab:dynamic}, showed that the test on \dynamicmodels different devices across \dynamicbrands brands identified \dynamicpropertiescase vulnerable system properties and \dynamicsettingscase vulnerable system settings among them.
This further confirmed our findings that the same vulnerabilities do exist across different devices from the same brand. We speculate that this may be due to code reuse within the same brand or among upstream suppliers.
The most prevalent vulnerable system settings were still ``xxx\_xxx\_i'' and ``xxx.xxx.deviceid'', both introduced by Baidu.
After decrypting their content, we found that on devices with higher Android versions (higher than 11), the Baidu SDK typically cannot obtain the device's IMEI. 
We speculate that on these newer versions, the Baidu SDK primarily relies on device identifiers generated from the ``android\_id'' for user tracking.
By conducting these dynamic tests, we demonstrated that similar vulnerable system properties and settings issues indeed exist across devices of the same brand.

We have timely disclosed the vulnerabilities identified in dynamic testing to vendors.
In general, vendors require corresponding evidences (e.g., PoCs) to confirm the vulnerabilities. Thus, we only reported the vulnerabilities that we have generated PoCs in real devices with system versions greater than 12, to the corresponding vendors.
We did not report issues in devices with older Android versions, because vendors typically focus on the security of recently released products, and issues present in lower versions may have already been fixed in higher versions.
Currently, two vendors have confirmed our findings, including 82 vulnerable system properties and 2 vulnerable settings. 
They have already fixed the related issues in the new system updates. 
The security vulnerability reports we sent to other vendors are still under review.

\noindent \textbf{Our Findings.}
\textit{We observed that vulnerable system properties/settings often repeatedly appear in devices from the same OEMs. 
By black-box testing of \dynamicmodels devices, we indeed found many recurring vulnerable properties and settings without ROMs.
We further reported our findings to respective vendors and received acknowledgment.}

\subsection{Ethical Consideration}
We have reported all our findings to the relevant parties and assisted them in addressing the issues. 
To date, all responding manufacturers have completed the issue fixes.
Additionally, we anonymized the brand names mentioned in the specific issue descriptions, as well as the detailed names of vulnerable system properties and settings in our paper.
There are two primary reasons. 
First, we are waiting for the remaining vendors to confirm our findings and address the security issues, followed by an additional 90-day confidentiality period, recommended by responsible disclosure. 
Second, and of greater concern, it is the fact that although vendors typically promptly address confirmed issues and strive to push security updates to users, a significant number of users continue to use older models that no longer receive system updates or may refuse to update their custom systems. 
These circumstances could result in custom Android systems on user devices containing the system properties or settings containing device non-resettable identifiers that we discovered.
Since our findings are derived from analyzing a large number of custom systems from various brands worldwide, if malicious developers exploit our discoveries to attempt to collect device non-resettable identifiers through the system properties and settings on user devices, it could pose a significant threat to user privacy.
Therefore, we chose to anonymize our specific findings to reduce the risk of compromising user privacy.

%% file: Sections/discussion.tex
\section{Discussion}
\label{sec:discussion}

\subsection{Mitigation}
To counteract the privacy risks posed by additional system properties and settings containing device non-resettable identifiers, we propose the following mitigation suggestions:

\noindent \textbf{More Explicit System Settings Access Control Methods.}
While Android has implemented relatively clear and comprehensive access control approaches for system properties through SELinux, the access control methods for system settings are much more ambiguous in comparison.
In the official Android documentation, there is far less content about system settings compared to system properties. 
We could not even find clear explanations for access control of system settings in the documentation, and the only related information we found was in the comments of the AOSP source code.
Therefore, we suggest that Google consider implementing additional access control mechanisms for system settings and tightening policies for newly added settings, so they are not accessible to third-party apps by default.

\noindent \textbf{Transparency of Covert Sensitive System Properties and Settings.}
While Google provides access control channels for different supply-chain actors (e.g., vendors and ODMs) in system customization when setting SELinux rules~\cite{vendorprop}, the number of actors involved in system customization far exceed these.
This results in access control enforcers may be unaware of the introduction of certain system properties or settings, such as the two sensitive system settings introduced by Baidu SDK.
We recommend that all actors involved in system customization provide detailed reports to access control enforcers when introducing sensitive system properties or settings. 
This would allow for comprehensive and effective access control to safeguard sensitive non-resettable device identifiers.

\noindent \textbf{Conducting Thorough Security Testing before Release.}
Upon verification, the ROMs of 1,516 devices in our dataset were confirmed to be supported by Google Play~\cite{playsupported}, indicating they passed Google's tests and should have undergone further testing by the device vendors before commercial deployment.
However, our findings show that testing for sensitive system properties and settings is inadequate.
We propose a simple yet effective testing method based on the characteristics of non-resettable identifiers to help reduce cases where sensitive properties or settings lack proper access control. 
This method involves two steps: First, after initializing the custom system, thoroughly use the device to ensure all sensitive properties and settings are initialized, then record their values. 
Second, factory reset the device and repeat the process, and identify properties and settings whose values remain unchanged.
Due to the nature of device non-resettable identifiers, which remain unchanged after a factory reset, this method can identify potential sensitive system properties and settings, and it can also handle cases where device identifiers are stored in encrypted form.

\subsection{Limitations}
\label{sec:limitations}

This paper has several limitations. 
First, we identified the system properties and settings present in custom Android systems by locating their usage, this approach, while addressing feasibility, also introduced constraints. 
Our approach does not allow us to accurately determine whether a system property or setting exists on the device, nor can it identify the specific circumstances under which they would be present.
Despite we have employed all possible methods to eliminate potential false positives, it is undeniable that some of them still persist.
Furthermore, our reliance on static analysis combined with the extensive code base of the custom system precludes us from obtaining the ground truth. 
This limitation greatly hinders our verification of whether all access behaviors of system properties and settings have been captured, and to ascertain if any sensitive ones have been missed.
Additionally, we only considered code written in Java language and did not take into account the access behaviors of system properties and settings in other languages, such as native code, which may have resulted in false negatives in our research.
Secondly, due to limitations in devices and financial resources, we could only use remote real-device testing services to conduct dynamic testing and validate our results. 
As mentioned in~\S\ref{sec:result}, it is difficult to find devices in remote real-device testing services that are exactly the same as those in our custom system dataset, making it impossible to generate PoCs for all the vulnerabilities we identified.
Additionally, the state of devices in testing services is complex, and we cannot confirm whether the devices have been fully initialized to reveal all sensitive system properties and settings. 
Moreover, these devices typically do not have SIM cards installed, which may result in the absence of sensitive system properties and settings related to SIM cards (e.g., IMSI, ICCID).

%% file: Sections/related.tex
\section{Related Work}
\label{sec:related}

\noindent \textbf{User Identifiers and Tracking Practices.}
Many existing researches have studied user identifiers and third-party tracking practices in mobile apps.
Razaghpanah et al.~\cite{razaghpanah2018apps} conducted a global study on mobile tracking ecosystems using the Lumen Privacy Monitor app. 
They identified 2,121 third-party advertising and tracking services, highlighting the heavy reliance on device identifiers for user tracking and the dominance of a few companies in the ecosystem.
Leith et al.~\cite{leith2021mobile} investigated data transmissions from mobile operating systems to manufacturers, revealing that device identifiers and other user data were sent to back-end servers even when users had configured their systems for minimal data sharing.
Kollnig et al.~\cite{kollnig2022iphones} studied 24,000 Android and iOS apps to compare the performance of these operating systems in terms of user privacy, noting the widespread use of user identifiers in both.
While these researchers examined the use of device identifiers for user tracking by Android apps and the system, they did not explore the use of identifiers beyond those provided by the documented system APIs.

\noindent \textbf{System Customization.}
Numerous previous studies have examined the security issues arising from the customizations from manufacturers.
Gamba et al.~\cite{gamba2020analysis} conducted a large-scale study of pre-installed apps on Android devices from over 200 vendors. 
They uncovered relationships between supply chain actors and found that these apps often exhibit invasive or even malicious behaviors, including backdoor access to sensitive data and services.
Lyons et al.~\cite{lyons2023log} systematically investigated the sensitive information stored in Android system logs. 
They discovered that many device identifiers and user behavior data were present in the logs, and that some high-privileged apps, including those from OEMs and large commercial companies, were collecting this information.
El-Rewini et al.~\cite{el2021dissecting} conducted a large-scale study on residual APIs—unused custom APIs left in the custom Android codebase. 
Analyzing 628 ROMs from seven major vendors, they found that these residual APIs are widespread and pose significant security risks, such as access control anomalies and potential exploitation.
These studies have explored specific types of issues introduced by system customizations, but none of them have involved the additional channels for obtaining device identifiers in custom systems.

Most relevant to our work, U2-I2~\cite{meng2023post} conducted a systematic study on system-level protection for non-resettable identifiers.
They revealed through small-scale dynamic analysis that in custom Android systems, many identifiers, including those in system properties and settings, are not adequately protected, potentially leading to leakage.
In addition to the seven well-known identifiers mentioned in~\S\ref{sec:background0}, they identified other non-resettable identifiers related to hardware (e.g., screen, NFC, and camera), which can remain constant after a factory reset and are similarly unprotected.
However, due to the limitations of dynamic methods, they are unable to perform large-scale analysis.
In contrast, our static approach requires only system ROMs, making it far more efficient and scalable. 

%% file: Sections/conclusion.tex
\section{Conclusion}
\label{sec:conclusion}
In this paper, we present a comprehensive analysis of non-resettable device identifiers in custom Android systems. We design an end-to-end approach to identify sensitive and vulnerable custom system properties and settings in the wild. 
By applying our approach to \datasetsize custom Android ROMs, we have identified thousands of vulnerable system properties and settings, and further validation through remote real-device testing confirmed our findings.
Through comparison, our approach outperforms existing work in scalability, efficiency and vulnerability coverage.
We also investigated the root reasons for such vulnerabilities, including overlooked system properties and settings, overly complex access control rules, and additional access channels for system properties. We also observed that the same vulnerable implementations usually recur across devices from the same brand.
Our study highlights the need for greater scrutiny of additional access channels to non-resettable device identifiers and the importance of better solutions to safeguard user privacy during system customizations.

%% file: Sections/appendices.tex
\appendix
\label{sec:appendix}
\section*{\centering \MakeUppercase{Appendices}}

\subsection*{SELinux Concepts}
Besides the concepts introduced in~\S\ref{sec:background2}, SELinux also provides the ``type attribute set'' mechanism for grouping related types under a single attribute, simplifying policy management and enforcement, as shown in example (e) of~\autoref{tab:selinux}.
By grouping types, systems can apply access control rules to multiple types simultaneously, ensuring consistency and reducing redundancy in policy definitions, as shown in example (d) of~\autoref{tab:selinux}.
It is worth mentioning that a type attribute set may consist of specific types and other sets, using logical operators ``and'' to include and ``not'' to exclude types. 
Additionally, the \texttt{expandtypeattribute} keyword determines if the members of a type attribute set are expanded (i.e., the specific types within the set are considered) during policy processing, as illustrated in example (f).
By default, type attribute sets are allowed to be expanded. 
Sets that typically do not require fine-grained control or those containing many types are often set to not allow expansion. 
Expanding sets that contain a large number of types during access control enforcement can consume a lot of computational overhead, leading to a slowdown in system performance.
In addition to regular policy rules, there is a class of rules in SELinux known as ``neverallow'' rules.
The neverallow rule is designed to specify actions that are strictly forbidden, even if other policies would allow them, helping prevent accidental or malicious policy configurations, and ensuring that critical security boundaries are not breached.
During the compilation phase, neverallow rules validate policy configurations. 
If a policy violates a neverallow rule, the compilation process will fail, preventing unsafe policies from being applied.
However, neverallow rules are enforced only during the compilation phase and are not dynamically checked at runtime.

\begin{table*}[t]
\caption{Breakdown of Results by Identifier Type.}
\label{tab:typesplit}
\begin{threeparttable}
\resizebox{1\textwidth}{!}{
\begin{tabular}{|c|ccccccc|ccccccc|}
\hline
\multirow{2}{*}{Brand} & \multicolumn{7}{c|}{System Property}                                                                                                                                                                  & \multicolumn{7}{c|}{System Setting}                                                                                                                                                            \\ \cline{2-15} 
                       & \multicolumn{1}{c|}{IMEI}     & \multicolumn{1}{c|}{IMSI}   & \multicolumn{1}{c|}{MEID}    & \multicolumn{1}{c|}{ICCID}    & \multicolumn{1}{c|}{SN}        & \multicolumn{1}{c|}{WiFi Mac} & BT Mac  & \multicolumn{1}{c|}{IMEI}     & \multicolumn{1}{c|}{IMSI}    & \multicolumn{1}{c|}{MEID} & \multicolumn{1}{c|}{ICCID}   & \multicolumn{1}{c|}{SN}   & \multicolumn{1}{c|}{WiFi Mac} & BT Mac   \\ \hline
Alps                   & \multicolumn{1}{c|}{23/11}    & \multicolumn{1}{c|}{10/0}   & \multicolumn{1}{c|}{11/7}    & \multicolumn{1}{c|}{33/19}    & \multicolumn{1}{c|}{14/2}      & \multicolumn{1}{c|}{0/0}      & 3/1     & \multicolumn{1}{c|}{4/0}      & \multicolumn{1}{c|}{6/1}     & \multicolumn{1}{c|}{0/0}  & \multicolumn{1}{c|}{2/1}     & \multicolumn{1}{c|}{0/0}  & \multicolumn{1}{c|}{2/1}      & 10/2     \\ \hline
Asus                   & \multicolumn{1}{c|}{18/17}    & \multicolumn{1}{c|}{0/0}    & \multicolumn{1}{c|}{0/0}     & \multicolumn{1}{c|}{0/0}      & \multicolumn{1}{c|}{16/5}      & \multicolumn{1}{c|}{0/0}      & 2/0     & \multicolumn{1}{c|}{5/0}      & \multicolumn{1}{c|}{4/0}     & \multicolumn{1}{c|}{0/0}  & \multicolumn{1}{c|}{5/0}     & \multicolumn{1}{c|}{5/0}  & \multicolumn{1}{c|}{6/0}      & 10/3     \\ \hline
Digma                  & \multicolumn{1}{c|}{2/2}      & \multicolumn{1}{c|}{2/0}    & \multicolumn{1}{c|}{1/1}     & \multicolumn{1}{c|}{8/0}      & \multicolumn{1}{c|}{7/0}       & \multicolumn{1}{c|}{0/0}      & 0/0     & \multicolumn{1}{c|}{10/0}     & \multicolumn{1}{c|}{0/0}     & \multicolumn{1}{c|}{0/0}  & \multicolumn{1}{c|}{0/0}     & \multicolumn{1}{c|}{0/0}  & \multicolumn{1}{c|}{0/0}      & 25/0     \\ \hline
Huawei                 & \multicolumn{1}{c|}{0/0}      & \multicolumn{1}{c|}{0/0}    & \multicolumn{1}{c|}{0/0}     & \multicolumn{1}{c|}{2/0}      & \multicolumn{1}{c|}{4/1}       & \multicolumn{1}{c|}{0/0}      & 0/0     & \multicolumn{1}{c|}{2/0}      & \multicolumn{1}{c|}{0/0}     & \multicolumn{1}{c|}{0/0}  & \multicolumn{1}{c|}{0/0}     & \multicolumn{1}{c|}{0/0}  & \multicolumn{1}{c|}{0/0}      & 3/1      \\ \hline
Lenovo                 & \multicolumn{1}{c|}{56/52}    & \multicolumn{1}{c|}{34/0}   & \multicolumn{1}{c|}{25/19}   & \multicolumn{1}{c|}{45/15}    & \multicolumn{1}{c|}{341/225}   & \multicolumn{1}{c|}{10/10}    & 18/16   & \multicolumn{1}{c|}{99/21}    & \multicolumn{1}{c|}{2/0}     & \multicolumn{1}{c|}{0/0}  & \multicolumn{1}{c|}{0/0}     & \multicolumn{1}{c|}{0/0}  & \multicolumn{1}{c|}{13/1}     & 68/29    \\ \hline
Meizu                  & \multicolumn{1}{c|}{70/35}    & \multicolumn{1}{c|}{0/0}    & \multicolumn{1}{c|}{36/23}   & \multicolumn{1}{c|}{58/48}    & \multicolumn{1}{c|}{54/19}     & \multicolumn{1}{c|}{0/0}      & 0/0     & \multicolumn{1}{c|}{79/29}    & \multicolumn{1}{c|}{0/0}     & \multicolumn{1}{c|}{0/0}  & \multicolumn{1}{c|}{8/0}     & \multicolumn{1}{c|}{0/0}  & \multicolumn{1}{c|}{20/0}     & 35/15    \\ \hline
Motorola               & \multicolumn{1}{c|}{12/12}    & \multicolumn{1}{c|}{18/0}   & \multicolumn{1}{c|}{1/1}     & \multicolumn{1}{c|}{40/6}     & \multicolumn{1}{c|}{329/132}   & \multicolumn{1}{c|}{6/5}      & 10/6    & \multicolumn{1}{c|}{17/8}     & \multicolumn{1}{c|}{12/0}    & \multicolumn{1}{c|}{0/0}  & \multicolumn{1}{c|}{0/0}     & \multicolumn{1}{c|}{0/0}  & \multicolumn{1}{c|}{6/3}      & 111/44   \\ \hline
Nokia                  & \multicolumn{1}{c|}{71/71}    & \multicolumn{1}{c|}{18/0}   & \multicolumn{1}{c|}{56/56}   & \multicolumn{1}{c|}{147/129}  & \multicolumn{1}{c|}{69/8}      & \multicolumn{1}{c|}{2/2}      & 0/0     & \multicolumn{1}{c|}{20/8}     & \multicolumn{1}{c|}{7/3}     & \multicolumn{1}{c|}{0/0}  & \multicolumn{1}{c|}{10/1}    & \multicolumn{1}{c|}{0/0}  & \multicolumn{1}{c|}{14/1}     & 49/23    \\ \hline
Nubia                  & \multicolumn{1}{c|}{21/21}    & \multicolumn{1}{c|}{0/0}    & \multicolumn{1}{c|}{11/11}   & \multicolumn{1}{c|}{3/3}      & \multicolumn{1}{c|}{35/6}      & \multicolumn{1}{c|}{0/0}      & 8/5     & \multicolumn{1}{c|}{39/12}    & \multicolumn{1}{c|}{0/0}     & \multicolumn{1}{c|}{0/0}  & \multicolumn{1}{c|}{0/0}     & \multicolumn{1}{c|}{0/0}  & \multicolumn{1}{c|}{23/4}     & 25/9     \\ \hline
Oneplus                & \multicolumn{1}{c|}{0/0}      & \multicolumn{1}{c|}{8/0}    & \multicolumn{1}{c|}{0/0}     & \multicolumn{1}{c|}{48/40}    & \multicolumn{1}{c|}{139/66}    & \multicolumn{1}{c|}{0/0}      & 0/0     & \multicolumn{1}{c|}{2/0}      & \multicolumn{1}{c|}{0/0}     & \multicolumn{1}{c|}{0/0}  & \multicolumn{1}{c|}{16/16}   & \multicolumn{1}{c|}{2/2}  & \multicolumn{1}{c|}{21/2}     & 37/27    \\ \hline
Oppo                   & \multicolumn{1}{c|}{67/66}    & \multicolumn{1}{c|}{34/0}   & \multicolumn{1}{c|}{2/2}     & \multicolumn{1}{c|}{90/56}    & \multicolumn{1}{c|}{191/105}   & \multicolumn{1}{c|}{0/0}      & 0/0     & \multicolumn{1}{c|}{50/6}     & \multicolumn{1}{c|}{0/0}     & \multicolumn{1}{c|}{0/0}  & \multicolumn{1}{c|}{0/0}     & \multicolumn{1}{c|}{6/0}  & \multicolumn{1}{c|}{30/1}     & 59/18    \\ \hline
Poco                   & \multicolumn{1}{c|}{117/1}    & \multicolumn{1}{c|}{30/0}   & \multicolumn{1}{c|}{34/3}    & \multicolumn{1}{c|}{52/26}    & \multicolumn{1}{c|}{90/25}     & \multicolumn{1}{c|}{4/2}      & 36/12   & \multicolumn{1}{c|}{32/12}    & \multicolumn{1}{c|}{19/15}   & \multicolumn{1}{c|}{0/0}  & \multicolumn{1}{c|}{12/9}    & \multicolumn{1}{c|}{0/0}  & \multicolumn{1}{c|}{35/19}    & 22/14    \\ \hline
Qti                    & \multicolumn{1}{c|}{93/24}    & \multicolumn{1}{c|}{1/1}    & \multicolumn{1}{c|}{36/17}   & \multicolumn{1}{c|}{38/38}    & \multicolumn{1}{c|}{125/39}    & \multicolumn{1}{c|}{2/2}      & 31/11   & \multicolumn{1}{c|}{41/18}    & \multicolumn{1}{c|}{22/15}   & \multicolumn{1}{c|}{0/0}  & \multicolumn{1}{c|}{18/13}   & \multicolumn{1}{c|}{0/0}  & \multicolumn{1}{c|}{49/20}    & 43/27    \\ \hline
Realme                 & \multicolumn{1}{c|}{34/34}    & \multicolumn{1}{c|}{54/0}   & \multicolumn{1}{c|}{1/1}     & \multicolumn{1}{c|}{96/42}    & \multicolumn{1}{c|}{211/115}   & \multicolumn{1}{c|}{0/0}      & 0/0     & \multicolumn{1}{c|}{23/4}     & \multicolumn{1}{c|}{0/0}     & \multicolumn{1}{c|}{0/0}  & \multicolumn{1}{c|}{0/0}     & \multicolumn{1}{c|}{4/0}  & \multicolumn{1}{c|}{45/0}     & 65/33    \\ \hline
Redmi                  & \multicolumn{1}{c|}{186/12}   & \multicolumn{1}{c|}{66/0}   & \multicolumn{1}{c|}{59/13}   & \multicolumn{1}{c|}{48/12}    & \multicolumn{1}{c|}{212/48}    & \multicolumn{1}{c|}{12/10}    & 90/29   & \multicolumn{1}{c|}{22/10}    & \multicolumn{1}{c|}{37/25}   & \multicolumn{1}{c|}{0/0}  & \multicolumn{1}{c|}{32/22}   & \multicolumn{1}{c|}{0/0}  & \multicolumn{1}{c|}{73/30}    & 44/13    \\ \hline
Samsung                & \multicolumn{1}{c|}{47/32}    & \multicolumn{1}{c|}{0/0}    & \multicolumn{1}{c|}{25/16}   & \multicolumn{1}{c|}{17/13}    & \multicolumn{1}{c|}{424/146}   & \multicolumn{1}{c|}{0/0}      & 21/19   & \multicolumn{1}{c|}{162/13}   & \multicolumn{1}{c|}{151/56}  & \multicolumn{1}{c|}{0/0}  & \multicolumn{1}{c|}{0/0}     & \multicolumn{1}{c|}{0/0}  & \multicolumn{1}{c|}{78/78}    & 133/65   \\ \hline
Xiaomi                 & \multicolumn{1}{c|}{763/160}  & \multicolumn{1}{c|}{16/0}   & \multicolumn{1}{c|}{189/48}  & \multicolumn{1}{c|}{14/4}     & \multicolumn{1}{c|}{400/139}   & \multicolumn{1}{c|}{13/11}    & 203/79  & \multicolumn{1}{c|}{163/72}   & \multicolumn{1}{c|}{116/56}  & \multicolumn{1}{c|}{0/0}  & \multicolumn{1}{c|}{48/29}   & \multicolumn{1}{c|}{0/0}  & \multicolumn{1}{c|}{187/72}   & 133/52   \\ \hline
Zte                    & \multicolumn{1}{c|}{72/72}    & \multicolumn{1}{c|}{6/0}    & \multicolumn{1}{c|}{25/25}   & \multicolumn{1}{c|}{21/15}    & \multicolumn{1}{c|}{24/4}      & \multicolumn{1}{c|}{0/0}      & 5/0     & \multicolumn{1}{c|}{67/32}    & \multicolumn{1}{c|}{62/1}    & \multicolumn{1}{c|}{0/0}  & \multicolumn{1}{c|}{23/10}   & \multicolumn{1}{c|}{1/1}  & \multicolumn{1}{c|}{14/3}     & 21/10    \\ \hline
Other (232)            & \multicolumn{1}{c|}{402/359}  & \multicolumn{1}{c|}{248/49} & \multicolumn{1}{c|}{103/90}  & \multicolumn{1}{c|}{415/196}  & \multicolumn{1}{c|}{543/104}   & \multicolumn{1}{c|}{15/14}    & 49/20   & \multicolumn{1}{c|}{164/21}   & \multicolumn{1}{c|}{23/9}    & \multicolumn{1}{c|}{1/0}  & \multicolumn{1}{c|}{14/5}    & \multicolumn{1}{c|}{4/1}  & \multicolumn{1}{c|}{48/8}     & 390/151  \\ \hline
Total (250)            & \multicolumn{1}{c|}{2054/981} & \multicolumn{1}{c|}{545/50} & \multicolumn{1}{c|}{615/333} & \multicolumn{1}{c|}{1175/662} & \multicolumn{1}{c|}{3228/1189} & \multicolumn{1}{c|}{64/56}    & 476/198 & \multicolumn{1}{c|}{1001/266} & \multicolumn{1}{c|}{461/181} & \multicolumn{1}{c|}{1/0}  & \multicolumn{1}{c|}{188/106} & \multicolumn{1}{c|}{22/4} & \multicolumn{1}{c|}{664/243}  & 1283/536 \\ \hline
\end{tabular}
}
\begin{tablenotes}
	\footnotesize
	\item Note: In each cell of the table, we provide two numbers in the format "num1/num2", where num1 is the number of sensitive system properties or settings, num2 is the number \\of vulnerable system properties or settings.
\end{tablenotes}
\end{threeparttable}
\end{table*}

%% file: main.bbl

\begin{thebibliography}{54}


\ifx \showCODEN    \undefined \def \showCODEN     #1{\unskip}     \fi
\ifx \showDOI      \undefined \def \showDOI       #1{#1}\fi
\ifx \showISBNx    \undefined \def \showISBNx     #1{\unskip}     \fi
\ifx \showISBNxiii \undefined \def \showISBNxiii  #1{\unskip}     \fi
\ifx \showISSN     \undefined \def \showISSN      #1{\unskip}     \fi
\ifx \showLCCN     \undefined \def \showLCCN      #1{\unskip}     \fi
\ifx \shownote     \undefined \def \shownote      #1{#1}          \fi
\ifx \showarticletitle \undefined \def \showarticletitle #1{#1}   \fi
\ifx \showURL      \undefined \def \showURL       {\relax}        \fi
\providecommand\bibfield[2]{#2}
\providecommand\bibinfo[2]{#2}
\providecommand\natexlab[1]{#1}
\providecommand\showeprint[2][]{arXiv:#2}

\bibitem[gdp(2024a)]%
        {gdpr}
 \bibinfo{year}{2024}\natexlab{a}.
\newblock \bibinfo{howpublished}{\url{https://gdpr-info.eu/}}.
\newblock


\bibitem[ven(2024)]%
        {vendorprop}
 \bibinfo{year}{2024}\natexlab{}.
\newblock \bibinfo{title}{Adding vendor-specific properties}.
\newblock \bibinfo{howpublished}{\url{https://source.android.com/docs/core/architecture/configuration/add-system-properties\#add-vendor-properties}}.
\newblock


\bibitem[adv(2024)]%
        {advertisingid}
 \bibinfo{year}{2024}\natexlab{}.
\newblock \bibinfo{title}{Advertising ID}.
\newblock \bibinfo{howpublished}{\url{https://support.google.com/googleplay/android-developer/answer/6048248?hl=en}}.
\newblock


\bibitem[ali(2024a)]%
        {alibaba}
 \bibinfo{year}{2024}\natexlab{a}.
\newblock \bibinfo{title}{Alibaba}.
\newblock \bibinfo{howpublished}{\url{https://www.alibaba.com/}}.
\newblock


\bibitem[ali(2024b)]%
        {aliyun}
 \bibinfo{year}{2024}\natexlab{b}.
\newblock \bibinfo{title}{Alibaba Cloud}.
\newblock \bibinfo{howpublished}{\url{https://cn.aliyun.com/}}.
\newblock


\bibitem[and(2024a)]%
        {android_dumps}
 \bibinfo{year}{2024}\natexlab{a}.
\newblock \bibinfo{title}{Android Dumps}.
\newblock \bibinfo{howpublished}{\url{https://dumps.tadiphone.dev/dumps}}.
\newblock


\bibitem[aos(2024)]%
        {aosp}
 \bibinfo{year}{2024}\natexlab{}.
\newblock \bibinfo{title}{Android Open Source Project}.
\newblock \bibinfo{howpublished}{\url{https://source.android.com/}}.
\newblock


\bibitem[and(2024b)]%
        {androidid}
 \bibinfo{year}{2024}\natexlab{b}.
\newblock \bibinfo{title}{ANDROID\_ID}.
\newblock \bibinfo{howpublished}{\url{https://developer.android.com/reference/android/provider/Settings.Secure\#ANDROID\_ID}}.
\newblock


\bibitem[sys(2024a)]%
        {systemservice}
 \bibinfo{year}{2024}\natexlab{a}.
\newblock \bibinfo{title}{Architecture overview}.
\newblock \bibinfo{howpublished}{\url{https://source.android.com/docs/core/architecture}}.
\newblock


\bibitem[ope(2024)]%
        {opensource}
 \bibinfo{year}{2024}\natexlab{}.
\newblock \bibinfo{title}{Artifact Open Source}.
\newblock \bibinfo{howpublished}{\url{https://github.com/r-paper/Custom-ROMs-Covert-Channels}}.
\newblock


\bibitem[bai(2024)]%
        {baidu}
 \bibinfo{year}{2024}\natexlab{}.
\newblock \bibinfo{title}{Baidu}.
\newblock \bibinfo{howpublished}{\url{https://home.baidu.com/}}.
\newblock


\bibitem[bin(2024)]%
        {binder}
 \bibinfo{year}{2024}\natexlab{}.
\newblock \bibinfo{title}{Binder}.
\newblock \bibinfo{howpublished}{\url{https://developer.android.com/reference/android/os/Binder}}.
\newblock


\bibitem[ccp(2024)]%
        {ccpa}
 \bibinfo{year}{2024}\natexlab{}.
\newblock \bibinfo{title}{California Consumer Privacy Act}.
\newblock \bibinfo{howpublished}{\url{https://oag.ca.gov/privacy/ccpa}}.
\newblock


\bibitem[tre(2024)]%
        {treble}
 \bibinfo{year}{2024}\natexlab{}.
\newblock \bibinfo{title}{Here comes Treble: A modular base for Android}.
\newblock \bibinfo{howpublished}{\url{https://android-developers.googleblog.com/2017/05/here-comes-treble-modular-base-for.html}}.
\newblock


\bibitem[pla(2024)]%
        {playsupported}
 \bibinfo{year}{2024}\natexlab{}.
\newblock \bibinfo{title}{https://support.google.com/googleplay/answer/1727131?hl=en}.
\newblock \bibinfo{howpublished}{\url{https://support.google.com/googleplay/answer/1727131?hl=en}}.
\newblock


\bibitem[len(2024)]%
        {lenovo}
 \bibinfo{year}{2024}\natexlab{}.
\newblock \bibinfo{title}{Lenovo}.
\newblock \bibinfo{howpublished}{\url{https://www.lenovo.com/}}.
\newblock


\bibitem[mac(2024)]%
        {mac}
 \bibinfo{year}{2024}\natexlab{}.
\newblock \bibinfo{title}{Mandatory access control}.
\newblock \bibinfo{howpublished}{\url{https://source.android.com/docs/security/features/selinux/concepts\#mandatory\_access\_control}}.
\newblock


\bibitem[med(2024)]%
        {mediatek}
 \bibinfo{year}{2024}\natexlab{}.
\newblock \bibinfo{title}{MediaTek | Powering the Brands you Love | Incredible Inside}.
\newblock \bibinfo{howpublished}{\url{https://www.mediatek.com/}}.
\newblock


\bibitem[mtk(2024)]%
        {mtk}
 \bibinfo{year}{2024}\natexlab{}.
\newblock \bibinfo{title}{MediaTek | Powering the Brands you Love | Incredible Inside}.
\newblock \bibinfo{howpublished}{\url{https://www.mediatek.com/}}.
\newblock


\bibitem[opp(2024)]%
        {oppo}
 \bibinfo{year}{2024}\natexlab{}.
\newblock \bibinfo{title}{Oppo}.
\newblock \bibinfo{howpublished}{\url{https://www.oppo.com/en/}}.
\newblock


\bibitem[sel(2024)]%
        {selinux}
 \bibinfo{year}{2024}\natexlab{}.
\newblock \bibinfo{title}{Security-Enhanced Linux in Android}.
\newblock \bibinfo{howpublished}{\url{https://source.android.com/docs/security/features/selinux}}.
\newblock


\bibitem[set(2024)]%
        {setools}
 \bibinfo{year}{2024}\natexlab{}.
\newblock \bibinfo{title}{SETools: Policy analysis tools for SELinux}.
\newblock \bibinfo{howpublished}{\url{https://github.com/SELinuxProject/setools}}.
\newblock


\bibitem[sys(2024b)]%
        {system_settings}
 \bibinfo{year}{2024}\natexlab{b}.
\newblock \bibinfo{title}{Settings}.
\newblock \bibinfo{howpublished}{\url{https://developer.android.com/reference/android/provider/Settings}}.
\newblock


\bibitem[sys(2024c)]%
        {system_properties}
 \bibinfo{year}{2024}\natexlab{c}.
\newblock \bibinfo{title}{System properties}.
\newblock \bibinfo{howpublished}{\url{https://source.android.com/docs/core/architecture/configuration\#system-properties}}.
\newblock


\bibitem[gdp(2024b)]%
        {gdprartifact}
 \bibinfo{year}{2024}\natexlab{b}.
\newblock \bibinfo{title}{U2-I2}.
\newblock \bibinfo{howpublished}{\url{https://uq-trust-lab.github.io/u2i2/}}.
\newblock


\bibitem[wri(2024a)]%
        {write_secure_settings}
 \bibinfo{year}{2024}\natexlab{a}.
\newblock \bibinfo{title}{WRITE\_SECURE\_SETTINGS}.
\newblock \bibinfo{howpublished}{\url{https://developer.android.com/reference/android/Manifest.permission\#WRITE\_SECURE\_SETTINGS}}.
\newblock


\bibitem[wri(2024b)]%
        {write_settings}
 \bibinfo{year}{2024}\natexlab{b}.
\newblock \bibinfo{title}{WRITE\_SETTINGS}.
\newblock \bibinfo{howpublished}{\url{https://developer.android.com/reference/android/Manifest.permission\#WRITE\_SETTINGS}}.
\newblock


\bibitem[xia(2024)]%
        {xiaomi}
 \bibinfo{year}{2024}\natexlab{}.
\newblock \bibinfo{title}{Xiaomi}.
\newblock \bibinfo{howpublished}{\url{https://www.mi.com/global/}}.
\newblock


\bibitem[Binns et~al\mbox{.}(2018)]%
        {binns2018third}
\bibfield{author}{\bibinfo{person}{Reuben Binns}, \bibinfo{person}{Ulrik Lyngs}, \bibinfo{person}{Max Van~Kleek}, \bibinfo{person}{Jun Zhao}, \bibinfo{person}{Timothy Libert}, {and} \bibinfo{person}{Nigel Shadbolt}.} \bibinfo{year}{2018}\natexlab{}.
\newblock \showarticletitle{Third party tracking in the mobile ecosystem}. In \bibinfo{booktitle}{\emph{Proceedings of the 10th ACM Conference on Web Science}}. \bibinfo{pages}{23--31}.
\newblock


\bibitem[Bl{\'a}zquez et~al\mbox{.}(2021)]%
        {blazquez2021trouble}
\bibfield{author}{\bibinfo{person}{Eduardo Bl{\'a}zquez}, \bibinfo{person}{Sergio Pastrana}, \bibinfo{person}{{\'A}lvaro Feal}, \bibinfo{person}{Julien Gamba}, \bibinfo{person}{Platon Kotzias}, \bibinfo{person}{Narseo Vallina-Rodriguez}, {and} \bibinfo{person}{Juan Tapiador}.} \bibinfo{year}{2021}\natexlab{}.
\newblock \showarticletitle{Trouble over-the-air: An analysis of fota apps in the android ecosystem}. In \bibinfo{booktitle}{\emph{2021 IEEE Symposium on Security and Privacy (SP)}}. IEEE, \bibinfo{pages}{1606--1622}.
\newblock


\bibitem[Chen et~al\mbox{.}(2014)]%
        {chen2014information}
\bibfield{author}{\bibinfo{person}{Terence Chen}, \bibinfo{person}{Imdad Ullah}, \bibinfo{person}{Mohamed~Ali Kaafar}, {and} \bibinfo{person}{Roksana Boreli}.} \bibinfo{year}{2014}\natexlab{}.
\newblock \showarticletitle{Information leakage through mobile analytics services}. In \bibinfo{booktitle}{\emph{Proceedings of the 15th Workshop on Mobile Computing Systems and Applications}}. \bibinfo{pages}{1--6}.
\newblock


\bibitem[Demetriou et~al\mbox{.}(2016)]%
        {demetriou2016free}
\bibfield{author}{\bibinfo{person}{Soteris Demetriou}, \bibinfo{person}{Whitney Merrill}, \bibinfo{person}{Wei Yang}, \bibinfo{person}{Aston Zhang}, {and} \bibinfo{person}{Carl~A Gunter}.} \bibinfo{year}{2016}\natexlab{}.
\newblock \showarticletitle{Free for all! assessing user data exposure to advertising libraries on android.}. In \bibinfo{booktitle}{\emph{NDSS}}.
\newblock


\bibitem[El-Rewini and Aafer(2021)]%
        {el2021dissecting}
\bibfield{author}{\bibinfo{person}{Zeinab El-Rewini} {and} \bibinfo{person}{Yousra Aafer}.} \bibinfo{year}{2021}\natexlab{}.
\newblock \showarticletitle{Dissecting residual APIs in custom android ROMs}. In \bibinfo{booktitle}{\emph{Proceedings of the 2021 ACM SIGSAC Conference on Computer and Communications Security}}. \bibinfo{pages}{1598--1611}.
\newblock


\bibitem[Elsabagh et~al\mbox{.}(2020)]%
        {elsabagh2020firmscope}
\bibfield{author}{\bibinfo{person}{Mohamed Elsabagh}, \bibinfo{person}{Ryan Johnson}, \bibinfo{person}{Angelos Stavrou}, \bibinfo{person}{Chaoshun Zuo}, \bibinfo{person}{Qingchuan Zhao}, {and} \bibinfo{person}{Zhiqiang Lin}.} \bibinfo{year}{2020}\natexlab{}.
\newblock \showarticletitle{$\{$FIRMSCOPE$\}$: Automatic uncovering of $\{$Privilege-Escalation$\}$ vulnerabilities in $\{$Pre-Installed$\}$ apps in android firmware}. In \bibinfo{booktitle}{\emph{29th USENIX security symposium (USENIX Security 20)}}. \bibinfo{pages}{2379--2396}.
\newblock


\bibitem[Gamba et~al\mbox{.}(2020)]%
        {gamba2020analysis}
\bibfield{author}{\bibinfo{person}{Julien Gamba}, \bibinfo{person}{Mohammed Rashed}, \bibinfo{person}{Abbas Razaghpanah}, \bibinfo{person}{Juan Tapiador}, {and} \bibinfo{person}{Narseo Vallina-Rodriguez}.} \bibinfo{year}{2020}\natexlab{}.
\newblock \showarticletitle{An analysis of pre-installed android software}. In \bibinfo{booktitle}{\emph{2020 IEEE symposium on security and privacy (SP)}}. IEEE, \bibinfo{pages}{1039--1055}.
\newblock


\bibitem[Hou et~al\mbox{.}(2022)]%
        {hou2022large}
\bibfield{author}{\bibinfo{person}{Qinsheng Hou}, \bibinfo{person}{Wenrui Diao}, \bibinfo{person}{Yanhao Wang}, \bibinfo{person}{Xiaofeng Liu}, \bibinfo{person}{Song Liu}, \bibinfo{person}{Lingyun Ying}, \bibinfo{person}{Shanqing Guo}, \bibinfo{person}{Yuanzhi Li}, \bibinfo{person}{Meining Nie}, {and} \bibinfo{person}{Haixin Duan}.} \bibinfo{year}{2022}\natexlab{}.
\newblock \showarticletitle{Large-scale security measurements on the android firmware ecosystem}. In \bibinfo{booktitle}{\emph{Proceedings of the 44th International Conference on Software Engineering}}. \bibinfo{pages}{1257--1268}.
\newblock


\bibitem[Kollnig et~al\mbox{.}(2022)]%
        {kollnig2022iphones}
\bibfield{author}{\bibinfo{person}{Konrad Kollnig}, \bibinfo{person}{Anastasia Shuba}, \bibinfo{person}{Reuben Binns}, \bibinfo{person}{Max Van~Kleek}, {and} \bibinfo{person}{Nigel Shadbolt}.} \bibinfo{year}{2022}\natexlab{}.
\newblock \showarticletitle{Are iphones really better for privacy? a comparative study of ios and android apps}.
\newblock \bibinfo{journal}{\emph{Proceedings on Privacy Enhancing Technologies}} \bibinfo{volume}{2022}, \bibinfo{number}{2} (\bibinfo{year}{2022}), \bibinfo{pages}{6--24}.
\newblock


\bibitem[Leith(2021)]%
        {leith2021mobile}
\bibfield{author}{\bibinfo{person}{Douglas~J Leith}.} \bibinfo{year}{2021}\natexlab{}.
\newblock \showarticletitle{Mobile handset privacy: Measuring the data iOS and Android send to Apple and Google}. In \bibinfo{booktitle}{\emph{Security and Privacy in Communication Networks: 17th EAI International Conference, SecureComm 2021, Virtual Event, September 6--9, 2021, Proceedings, Part II 17}}. Springer, \bibinfo{pages}{231--251}.
\newblock


\bibitem[Lerner et~al\mbox{.}(2016)]%
        {lerner2016internet}
\bibfield{author}{\bibinfo{person}{Ada Lerner}, \bibinfo{person}{Anna~Kornfeld Simpson}, \bibinfo{person}{Tadayoshi Kohno}, {and} \bibinfo{person}{Franziska Roesner}.} \bibinfo{year}{2016}\natexlab{}.
\newblock \showarticletitle{Internet jones and the raiders of the lost trackers: An archaeological study of web tracking from 1996 to 2016}. In \bibinfo{booktitle}{\emph{25th USENIX Security Symposium (USENIX Security 16)}}.
\newblock


\bibitem[Li et~al\mbox{.}(2021)]%
        {li2021android}
\bibfield{author}{\bibinfo{person}{Rui Li}, \bibinfo{person}{Wenrui Diao}, \bibinfo{person}{Zhou Li}, \bibinfo{person}{Jianqi Du}, {and} \bibinfo{person}{Shanqing Guo}.} \bibinfo{year}{2021}\natexlab{}.
\newblock \showarticletitle{Android custom permissions demystified: From privilege escalation to design shortcomings}. In \bibinfo{booktitle}{\emph{2021 IEEE Symposium on Security and Privacy (SP)}}. IEEE, \bibinfo{pages}{70--86}.
\newblock


\bibitem[Liu et~al\mbox{.}(2021)]%
        {liu2021android}
\bibfield{author}{\bibinfo{person}{Haoyu Liu}, \bibinfo{person}{Paul Patras}, {and} \bibinfo{person}{Douglas~J Leith}.} \bibinfo{year}{2021}\natexlab{}.
\newblock \showarticletitle{Android mobile OS snooping by Samsung, Xiaomi, Huawei and Realme handsets}.
\newblock \bibinfo{journal}{\emph{techreport, Oct}} (\bibinfo{year}{2021}).
\newblock


\bibitem[Liu et~al\mbox{.}(2022)]%
        {liu2022customized}
\bibfield{author}{\bibinfo{person}{Pei Liu}, \bibinfo{person}{Mattia Fazzini}, \bibinfo{person}{John Grundy}, {and} \bibinfo{person}{Li Li}.} \bibinfo{year}{2022}\natexlab{}.
\newblock \showarticletitle{Do customized Android frameworks keep pace with Android?}. In \bibinfo{booktitle}{\emph{Proceedings of the 19th International Conference on Mining Software Repositories}}. \bibinfo{pages}{376--387}.
\newblock


\bibitem[Lyons et~al\mbox{.}(2023)]%
        {lyons2023log}
\bibfield{author}{\bibinfo{person}{Allan Lyons}, \bibinfo{person}{Julien Gamba}, \bibinfo{person}{Austin Shawaga}, \bibinfo{person}{Joel Reardon}, \bibinfo{person}{Juan Tapiador}, \bibinfo{person}{Serge Egelman}, {and} \bibinfo{person}{Narseo Vallina-Rodr{\'\i}guez}.} \bibinfo{year}{2023}\natexlab{}.
\newblock \showarticletitle{Log:$\{$It’s$\}$ Big,$\{$It’s$\}$ Heavy,$\{$It’s$\}$ Filled with Personal Data! Measuring the Logging of Sensitive Information in the Android Ecosystem}. In \bibinfo{booktitle}{\emph{32nd USENIX Security Symposium (USENIX Security 23)}}. \bibinfo{pages}{2115--2132}.
\newblock


\bibitem[Meng et~al\mbox{.}(2023)]%
        {meng2023post}
\bibfield{author}{\bibinfo{person}{Mark~Huasong Meng}, \bibinfo{person}{Qing Zhang}, \bibinfo{person}{Guangshuai Xia}, \bibinfo{person}{Yuwei Zheng}, \bibinfo{person}{Yanjun Zhang}, \bibinfo{person}{Guangdong Bai}, \bibinfo{person}{Zhi Liu}, \bibinfo{person}{Sin~G Teo}, {and} \bibinfo{person}{Jin~Song Dong}.} \bibinfo{year}{2023}\natexlab{}.
\newblock \showarticletitle{Post-GDPR Threat Hunting on Android Phones: Dissecting OS-level Safeguards of User-unresettable Identifiers.}. In \bibinfo{booktitle}{\emph{NDSS}}.
\newblock


\bibitem[Nath(2015)]%
        {nath2015madscope}
\bibfield{author}{\bibinfo{person}{Suman Nath}.} \bibinfo{year}{2015}\natexlab{}.
\newblock \showarticletitle{Madscope: Characterizing mobile in-app targeted ads}. In \bibinfo{booktitle}{\emph{Proceedings of the 13th Annual International Conference on Mobile Systems, Applications, and Services}}. \bibinfo{pages}{59--73}.
\newblock


\bibitem[Possemato et~al\mbox{.}(2021)]%
        {possemato2021trust}
\bibfield{author}{\bibinfo{person}{Andrea Possemato}, \bibinfo{person}{Simone Aonzo}, \bibinfo{person}{Davide Balzarotti}, {and} \bibinfo{person}{Yanick Fratantonio}.} \bibinfo{year}{2021}\natexlab{}.
\newblock \showarticletitle{Trust, but verify: A longitudinal analysis of android oem compliance and customization}. In \bibinfo{booktitle}{\emph{2021 IEEE symposium on security and privacy (SP)}}. IEEE, \bibinfo{pages}{87--102}.
\newblock


\bibitem[Razaghpanah et~al\mbox{.}(2018)]%
        {razaghpanah2018apps}
\bibfield{author}{\bibinfo{person}{Abbas Razaghpanah}, \bibinfo{person}{Rishab Nithyanand}, \bibinfo{person}{Narseo Vallina-Rodriguez}, \bibinfo{person}{Srikanth Sundaresan}, \bibinfo{person}{Mark Allman}, \bibinfo{person}{Christian Kreibich}, \bibinfo{person}{Phillipa Gill}, {et~al\mbox{.}}} \bibinfo{year}{2018}\natexlab{}.
\newblock \showarticletitle{Apps, trackers, privacy, and regulators: A global study of the mobile tracking ecosystem}. In \bibinfo{booktitle}{\emph{The 25th Annual Network and Distributed System Security Symposium (NDSS 2018)}}.
\newblock


\bibitem[Reardon et~al\mbox{.}(2019)]%
        {reardon201950}
\bibfield{author}{\bibinfo{person}{Joel Reardon}, \bibinfo{person}{{\'A}lvaro Feal}, \bibinfo{person}{Primal Wijesekera}, \bibinfo{person}{Amit Elazari~Bar On}, \bibinfo{person}{Narseo Vallina-Rodriguez}, {and} \bibinfo{person}{Serge Egelman}.} \bibinfo{year}{2019}\natexlab{}.
\newblock \showarticletitle{50 ways to leak your data: An exploration of apps' circumvention of the android permissions system}. In \bibinfo{booktitle}{\emph{28th USENIX security symposium (USENIX security 19)}}. \bibinfo{pages}{603--620}.
\newblock


\bibitem[Ren et~al\mbox{.}(2018)]%
        {ren2018longitudinal}
\bibfield{author}{\bibinfo{person}{Jingjing Ren}, \bibinfo{person}{Martina Lindorfer}, \bibinfo{person}{Daniel~J Dubois}, \bibinfo{person}{Ashwin Rao}, \bibinfo{person}{David Choffnes}, {and} \bibinfo{person}{Narseo Vallina-Rodriguez}.} \bibinfo{year}{2018}\natexlab{}.
\newblock \showarticletitle{A longitudinal study of pii leaks across android app versions}. In \bibinfo{booktitle}{\emph{Network and Distributed System Security Symposium (NDSS)}}, Vol.~\bibinfo{volume}{10}.
\newblock


\bibitem[Schindler et~al\mbox{.}(2022)]%
        {schindler2022privacy}
\bibfield{author}{\bibinfo{person}{Christian Schindler}, \bibinfo{person}{M{\"u}sl{\"u}m Atas}, \bibinfo{person}{Thomas Strametz}, \bibinfo{person}{Johannes Feiner}, {and} \bibinfo{person}{Reinhard Hofer}.} \bibinfo{year}{2022}\natexlab{}.
\newblock \showarticletitle{Privacy leak identification in third-party android libraries}. In \bibinfo{booktitle}{\emph{2022 seventh international conference on mobile and secure services (MobiSecServ)}}. IEEE, \bibinfo{pages}{1--6}.
\newblock


\bibitem[Stevens et~al\mbox{.}(2012)]%
        {stevens2012investigating}
\bibfield{author}{\bibinfo{person}{Ryan Stevens}, \bibinfo{person}{Clint Gibler}, \bibinfo{person}{Jon Crussell}, \bibinfo{person}{Jeremy Erickson}, {and} \bibinfo{person}{Hao Chen}.} \bibinfo{year}{2012}\natexlab{}.
\newblock \showarticletitle{Investigating user privacy in android ad libraries}. In \bibinfo{booktitle}{\emph{Workshop on Mobile Security Technologies (MoST)}}, Vol.~\bibinfo{volume}{10}. \bibinfo{pages}{195--197}.
\newblock


\bibitem[Vall{\'e}e-Rai et~al\mbox{.}(2010)]%
        {vallee2010soot}
\bibfield{author}{\bibinfo{person}{Raja Vall{\'e}e-Rai}, \bibinfo{person}{Phong Co}, \bibinfo{person}{Etienne Gagnon}, \bibinfo{person}{Laurie Hendren}, \bibinfo{person}{Patrick Lam}, {and} \bibinfo{person}{Vijay Sundaresan}.} \bibinfo{year}{2010}\natexlab{}.
\newblock \showarticletitle{Soot: A Java bytecode optimization framework}.
\newblock In \bibinfo{booktitle}{\emph{CASCON First Decade High Impact Papers}}. \bibinfo{pages}{214--224}.
\newblock


\bibitem[Yang and Yue(2020)]%
        {yang2020comparative}
\bibfield{author}{\bibinfo{person}{Zhiju Yang} {and} \bibinfo{person}{Chuan Yue}.} \bibinfo{year}{2020}\natexlab{}.
\newblock \showarticletitle{A comparative measurement study of web tracking on mobile and desktop environments}.
\newblock \bibinfo{journal}{\emph{Proceedings on Privacy Enhancing Technologies}} \bibinfo{volume}{2020}, \bibinfo{number}{2} (\bibinfo{year}{2020}).
\newblock


\bibitem[Zhou et~al\mbox{.}(2022)]%
        {zhou2022uncovering}
\bibfield{author}{\bibinfo{person}{Hao Zhou}, \bibinfo{person}{Xiapu Luo}, \bibinfo{person}{Haoyu Wang}, {and} \bibinfo{person}{Haipeng Cai}.} \bibinfo{year}{2022}\natexlab{}.
\newblock \showarticletitle{Uncovering Intent based Leak of Sensitive Data in Android Framework}. In \bibinfo{booktitle}{\emph{Proceedings of the 2022 ACM SIGSAC Conference on Computer and Communications Security}}. \bibinfo{pages}{3239--3252}.
\newblock


\end{thebibliography}
